\documentclass{IEEEtran}

\usepackage{graphics,graphicx} 
\usepackage{epsfig} 

\usepackage{amsmath}
\usepackage{amssymb}
\usepackage{tikz}
\usepackage{amsthm}
\usepackage{comment}
\usepackage[export]{adjustbox}
\usepackage{mathcomSTEP}
\usepackage{subcaption,setspace}
\usepackage[utf8]{inputenc}

\usepackage{tikz}
\usetikzlibrary{positioning}
\usetikzlibrary{arrows,fit}

\tikzstyle{int}=[draw, fill=blue!10, minimum height = 1cm, minimum width=1.5cm,thick ]
\tikzstyle{joint} = [draw, circle, minimum size=1em]

\usepackage{epstopdf}
\usepackage{pgfplots,setspace}

\tikzstyle{int}=[draw, fill=blue!10, minimum height = 1cm, minimum width=1.5cm,thick ]
\tikzstyle{sum}=[circle, fill=black!10, draw=black,line width=1pt,minimum size = 0.3cm, thick ]
\tikzset{cross/.style={cross out, draw=black, minimum size=2*(#1-\pgflinewidth), inner sep=0pt, outer sep=0pt},
cross/.default={1pt}
}
\usetikzlibrary{shapes.misc}

\allowdisplaybreaks

\theoremstyle{remark}
\newtheorem{rem}[thm]{Remark}

\newtheorem{definition}{Definition}

\begin{document}

\title{
On the Capacity of \\the Carbon Copy onto Dirty Paper Channel
}

\author{%
\IEEEauthorblockN{%
Stefano Rini \IEEEauthorrefmark{1} and Shlomo Shamai (Shitz)\IEEEauthorrefmark{2} \\}

\IEEEauthorblockA{%
\IEEEauthorrefmark{1}
National Chiao-Tung University, Hsinchu, Taiwan\\
E-mail: \texttt{stefano@nctu.edu.tw} }

\IEEEauthorblockA{%
\IEEEauthorrefmark{2}
Technion-Israel Institute of Technology,  Haifa, Israel \\
E-mail: \texttt{sshlomo@ee.technion.ac.il}
}

\thanks{
The work of S. Rini was funded by the  Ministry Of Science and Technology (MOST) under the grant 103-2218-E-009-014-MY2.
The work of S. Shamai was supported by the European FP7 NEWCOM\#,
and the Heron Consortium 5G Technologies, Israel Ministry of Science.
Part of this work has been presented at the 2014 Information Theory Workshop (ITW), Hobart, Australia
and at the 2016  International Conference on the Science of Electrical
Engineering (ICSEE), Eilat, Israel.
}
}

\maketitle

\begin{abstract}
The ``Carbon Copy onto Dirty Paper'' (CCDP) channel is the compound ``writing on dirty paper'' channel in which the channel output is obtained as the sum of the channel input, white Gaussian noise and a Gaussian state sequence randomly selected among a set possible realizations.
The transmitter has non-causal knowledge of the set of possible state sequences but does not know which sequence is selected to produce the channel output.
%
%
We study the capacity of the CCDP  channel
for two scenarios: (i) the  state sequences  are independent and identically distributed,
and (ii)  the state sequences are scaled versions of the same sequence.
In the first scenario, we show that a combination of superposition coding, time-sharing and Gel'fand-Pinsker binning is sufficient
to approach the capacity to within three bits per channel use for any number of possible state realizations.
In the second scenario, we derive capacity to within four bits--per--channel--use for the  case of two possible state sequences.
This result is extended to the CCDP channel with any number of possible state sequences under certain conditions on the scaling parameters which we denote as ``strong fading'' regime.
We conclude by providing some remarks on the capacity of the CCDP channel in which the  state sequences have any jointly Gaussian distribution.
\end{abstract}


\section*{Introduction}
\label{sec:Introduction}

The Gel'fand-Pinsker (GP) channel \cite{GelfandPinskerClassic} is  the point-to-point channel in which the channel output is obtained as a random function of the  input and a state sequence which is provided non-causally to the encoder but is unknown at the decoder.
Costa's ``Writing on Dirty Paper'' (WDP) channel \cite{costa1983writing} is the Gaussian version of the GP channel in which the channel output is equal to the sum of the input, a channel state and white Gaussian noise.
In \cite{costa1983writing} Costa  proved that the transmitter can fully pre-code its transmissions against the channel state so that the capacity of the WDP channel is equal to the capacity of the Gaussian point-to-point channel.
Unfortunately, the performance of the capacity-achieving transmission scheme in \cite{costa1983writing} quickly degrades in the presence of  uncertainty in the channel knowledge:
for this reason, it is of great interest to extend Costa's result  to models in which only partial channel knowledge is available at the users.

In the following, we investigate the compound version of the WDP channel, the CCDP channel \cite{LapidothCarbonCopying}.
This channel models the WDP channel in which the channel state sequence is randomly drawn among a set possible realizations, all anti-causally known only at the transmitter.
The CCDP channel is obtained from the compound channel model \cite{lapidoth1998compound} by letting the output at each compound receiver equal the sum of the channel input, white Gaussian noise and a Gaussian state known only at the transmitter.
The CCDP channel is also equivalent to a Gaussian broadcast channel with a common message and with channel states known only at the transmitter \cite{steinberg2005achievable}.
%


\subsubsection*{Related Results}
%
%
The compound GP channel is the  discrete memoryless compound channel in which the output at each compound receiver is a random function
of the channel input and a state sequence non-causally known at  the  encoder.
An achievable region for the  two-receiver compound GP channel is presented in  \cite{piantanida2010capacity,nair2010achievability} where it is shown that using a common message improves over the coding scheme in which the transmitter simultaneously bins against both state realizations\footnote{Note that the capacity for this model was incorrectly claimed in \cite{piantanida2009capacity,moulin2007capacity}.}.
%
%
In \cite{LapidothCarbonCopying} the authors introduce  the CCDP channel as the compound GP channel with additive Gaussian state and additive Gaussian noise and derive the  first inner and outer bounds for to capacity.
%
%

The CCDP channel can be used to model the WDP channel affected by  slow fading  and with receiver side information.
This is obtained by letting the channel states be a scaled version of the same state sequence:  we term this model ``Writing on Slow Fading Dirt'' (WSFD) channel.
The fast fading counterpart of the WSFD channel, in which the state is multiplied by a fast fading process, is known as the ``Writing on Fast Fading Dirt'' (WFFD) channel\footnote{This model is also known as ``writing on faded dirt'' channel, ``fading dirty paper'' channel or ``dirty paper channel with fading dirt''.}.
The WFFD channel was first studied in  \cite{grover2007need}  for the case in which an \emph{i.i.d.} phase fading process affects the channel state.
In \cite{grover2007writing}, the same authors derive upper and lower bounds to the outage probability for this model.
Achievable rates under Gaussian signaling are derived in \cite{avner2010dirty} for the channel in which the state is multiplied by a Gaussian fast fading process.
The authors of \cite{zhang2007writing} consider the case in which  both the input and the state sequences are multiplied by the
fading process.
For this model, it is shown that the rate loss from full state pre-cancellation is  vanishing in both the ergodic and quasi-static fading case and at both high and low SNRs.
This result holds because fading affects the sum of state and input and thus Costa pre-coding as in the WDP channel is still effective.
The model above is further investigated in \cite{vaze2008dirty}, which also considers the multi-antenna setting.
In \cite{vaze2008dirty} algorithms are also proposed to determine the optimal linear pre-coding strategies which are shown to outperform Costa's linear assignment  in the multiple antenna setting.
%
%
In \cite{RiniPhase14}, we derive the capacity to within a constant gap  for the channel in which the fading only takes two possible values: this result is extended in \cite{rini2015dirty} to include  more general  fading distribution and to consider the case in which the fading sequence is not known either at the transmitter or at the receiver.

A model which encompasses the CCDP channel as a special case is the state-dependent broadcast channel with a common message.
This model  is obtained from the CCDP channel by introducing an additional private message to be communicated between the transmitter and each receiver.
A first achievable region for this channel is derived in \cite{steinberg2005coding} by combining coding strategies for the GP channel and the Gaussian broadcast channel.
The approximate capacity for the case of two receivers is determined in \cite{ghabeli2016capacity}.
The authors of \cite{biglieri2007coding} point out how the study of the Gaussian state-dependent broadcast channel with a common message appears more arduous than the study of the state-dependent broadcast channel with independent messages. This is due to the fact that the former model
is not degraded and thus the capacity region with a common message cannot be directly deduced from the capacity region of the channel with independent messages.

\subsubsection*{Contributions}
\label{sec:Contributions}
In the following, we investigate the capacity of the $M$-receiver CCDP channel.
We focus, in particular, on two classes which we term
(i) the ``Writing on Random Dirty Paper'' (WRDP)  channel and (ii) the ``Writing on Slow Fading Dirt'' (WSFD) channel.
The WRDP channel corresponds to the CCDP channel with \emph{i.i.d.} channel state sequences while the WSFD channel is the CCDP channel in which the state sequences are scaled versions of the same  sequence.
We also  consider a third model: (iii) the CCDP  with ``Equivalent States'' (CCDP-ES) channel in which the channel states have the same variance and the same pairwise correlation.
%
%
%

For the  models above, we characterize the approximate capacity\footnote{In the following, for brevity, we use the term  ``approximate capacity'' in lieu of ``capacity to within a constant gap''. A precise definition of ``approximate capacity'' is provided in Def. \ref{def:approximate}.} in the following classes:

\smallskip
\noindent
{\bf Sec. \ref{sec:The Writing on Random Dirty Paper (WRDP)}--
 WRDP channel:}
  For this model, we determine the approximate capacity for all parameter regimes and any number of compound receivers; we begin by considering the case of $M=2$  receivers and successively extend this result to any value of $M$.
  Capacity is approached by having the transmitter send the superposition of two codewords: the bottom codeword treats the state as noise and is decoded by all the users.
  The top codeword, instead,  is time-shared among all receivers as it is pre-coded against the state in the $m^{th}$ channel output for a portion $1/M$ of the time.
  %

  %
 \smallskip
\noindent
{\bf  Sec. \ref{sec:Dirty Paper Channel  with Slow Fading Dirt}--
  WSFD channel:}
  For this channel, we determine the approximate capacity for the case $M=2$ and generalize this result to the case any value $M$ only under some additional conditions on the channel parameters which we term ``strong fading'' conditions.
  %
  %
  As for the WRDP channel, the achievable strategies rely on superposition coding and state pre-cancellation with time-sharing among the different receivers.
  In the WSFD channel, though, simultaneous state pre-cancellation at multiple receivers is also necessary when channel states have high correlation.
  %

  \smallskip
\noindent
  {\bf Sec. \ref{sec:The general  Carbon Copy onto Dirty Paper (CCDP) channel}--
  CCDP-ES channel:}
  Here, as in the previous sections, we first derive the approximate capacity for the case of $M=2$   compound receivers and then generalize this result to any value of $M$.
  For  the CCDP-ES, we show that the channel state sequences can be decomposed in a common part, as in the  WFD channel, and in an independent part,
  as in the WRDP channel, so that a combination of the results in Sec.  \ref{sec:Dirty Paper Channel  with Slow Fading Dirt}
  and Sec. \ref{sec:Dirty Paper Channel  with Slow Fading Dirt} are sufficient to approach capacity.
  %
%

\subsection*{Paper Organization}
\label{sec:Paper Organization}

The remainder of the paper is organized as follows: in Sec. \ref{sec:Channel Model} we introduce the
CCDP
 channel and specialize this model
 to obtain  the WRDP, the
  WSFD,
    and the
   CCDP-ES channels.
%
Sec. \ref{sec:Related Results} presents the relevant results available in the literature.
In Sec.  \ref{sec:The Writing on Random Dirty Paper (WRDP)} we study the WRDP channel while, in Sec. \ref{sec:Dirty Paper Channel  with Slow Fading Dirt}, we investigate the
WSFD channel.
The CCDP-ES is considered in Sec. \ref{sec:The general  Carbon Copy onto Dirty Paper (CCDP) channel}.
Finally, Sec. \ref{sec:Conclusion} concludes the paper.
\section{Channel Model}
\label{sec:Channel Model}

\begin{figure}
\centering
\begin{tikzpicture}[node distance=2cm,auto,>=latex]
  \node at (0,0) (origin) {};
  \node at (-1,0) (source) {$W$};
  \node [int, right of = source,node distance = 1.4 cm](enc){Enc.};
  \node [sum, right of = enc,node distance = 1.7 cm](enc1){};
  \node [right of = enc1,node distance = 1cm](p11){+};
  \node [joint,right of = enc1,node distance = 1cm](p11){};
  \node [right of = p11,node distance =.75 cm](p12){+};
  \node [joint,right of = p11,node distance =.75 cm](p12){};
  \node [int, right of = p12,node distance = 1.8 cm](dec1){Dec. 2};
  \node [ right of = dec1,node distance = 1.75 cm](sink1){};
  \node [above of = p11,node distance = 1 cm](S1){$c S_2^N$};
  \node [above of = p12,node distance = 1 cm](Z1){$Z_2^N$};
  \draw[->,line width=1pt] (S1) -- (p11) ;
  \draw[->,line width=1pt] (Z1) -- (p12) ;
  \draw[-,line width=1pt] (p11) -- (p12) ;
  \draw[->,line width=1pt] (p12) node[above, xshift =0.7 cm] {$Y_2^N$}-- (dec1) ;
 \node [above of = p11,node distance = 2 cm](p21){+};
 \node [joint,above of = p11,node distance = 2 cm](p21){};
  \node [right of = p21,node distance =.75 cm](p22){+};
  \node [joint,right of = p21,node distance =.75 cm](p22){};
  \node [int, right of = p22,node distance = 1.8 cm](dec2){Dec. 1};
  \node [right of = dec2,node distance = 1.75 cm](sink2){};
  \node [above of = p21,node distance = 1 cm](S2){$c S_1^N$};
  \node [above of = p22,node distance = 1 cm](Z2){$Z_1^N$};
  \draw[->,line width=1pt] (S2) -- (p21) ;
  \draw[->,line width=1pt] (Z2) -- (p22) ;
  \draw[-,line width=1pt] (p21) -- (p22) ;
  \draw[->,line width=1pt] (p22) node[above, xshift =0.7 cm] {$Y_1^N$} -- (dec2) ;
  \node [below of = p11,node distance = 1 cm](p31){};
  \node [right of = p31,node distance = 1.5 cm](p32){$\vdots$};
  \node [below of = sink1,node distance = 1 cm](sink3){$\vdots$};
  \node [below of = dec1,node distance = 1 cm](dec3){$\vdots$};
  \node [below of = p31,node distance = 2 cm](p41){+};
  \node [joint,below of = p31,node distance = 2 cm](p41){};
  \node [right of = p41,node distance =.75 cm](p42){+};
  \node [joint,right of = p41,node distance =.75 cm](p42){};
  \node [int, right of = p42,node distance = 1.8 cm](dec4){Dec. M};
  \node [right of = dec4,node distance = 1.75 cm](sink4){};
 \draw[-,line width=1pt] (p41) -- (p42) ;
 \draw[->,line width=1pt] (p42) node[above, xshift =0.7 cm] {$Y_M^N$} -- (dec4) ;
 \node [above of = p41,node distance = 1 cm](SM){$c S_M^N$};
 \node [above of = p42,node distance = 1 cm](ZM){$Z_M^N$};
 \draw[->,line width=1pt] (SM) -- (p41) ;
 \draw[->,line width=1pt] (ZM) -- (p42) ;
 %
 %
 \draw[-,line width=1pt] (enc) node[above, xshift =1.25 cm] {$X^N$}--(enc1) {};
 \draw[->,line width=1pt, bend left=90 ]  (enc1) |- (p21);
 \draw[->,line width=1pt] (enc1) -- (p11);
 \draw[->,line width=1pt] (enc1) |- (p41);
 \draw[->,line width=.5 pt] (source) -- (enc);
 \draw[->,line width=.5 pt] (dec1) node[above, xshift =1.75 cm] {$\Wh(Y_2^N)$}--(sink1) ;
 \draw[->,line width=.5 pt] (dec2) node[above, xshift =1.75 cm] {$\Wh(Y_1^N)$} -- (sink2);
 \draw[->,line width=.5 pt] (dec4) node[above, xshift =1.75 cm] {$\Wh(Y_M^N)$}-- (sink4);
 \draw[dotted,->,line width=1pt]  (S1) -| (enc) ;
 \draw[->,line width=1pt,dotted]  (S2) -| (enc) ;
 \draw[->,line width=1pt,dotted]  (SM) -| (enc) ;
\%
\end{tikzpicture}
\caption{The ``Carbon Copying onto  Dirty Paper'' (CCDP) channel.}
\label{fig:CCDP channel}
\end{figure}
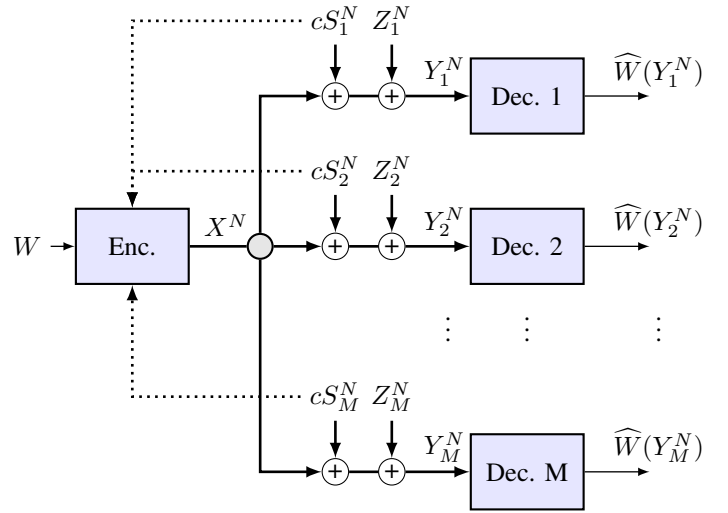

The $M$-receiver CCDP channel,  also depicted in Fig. \ref{fig:CCDP channel},  is the compound channel with states known at the transmitter
in which the  output at the $m^{\rm th}$ compound receiver is obtained as
\ea{
Y_m^N=X^N+c S_m^N+Z_m^N, \quad \quad m \in [1 \ldots M],
\label{eq:compoundChannel}
}
where $X^N$ is the channel input, $S_m^N$ the channel state sequence,  $Z_m^N$ a white Gaussian noise sequence with  zero mean and unitary variance and
$c \geq 0$ without loss of generality.
The transmitter, having knowledge of the state sequences, wishes to reliably communicate the message $W \in \Wcal=[1 \ldots 2^{NR}]$ to each of the $M$ compound receivers, despite the presence of the additive state and the additive noise.
The  channel input $X^N$  is  subject to the average power constraint
\ea{
 \sum_{i=1}^N \Ebb \lsb |X_i|^2 \rsb \leq N P.
 \label{eq:power constraint}    
}
For each channel use $i$, $[S_{1i} \ldots S_{Mi}]$ is an \emph{i.i.d.} jointly Gaussian random vector
with zero mean and covariance matrix $\Sigma_S$\footnote{In the following, we use the short-hand notation $[S_{1}^N \ldots S_{M}^N] \sim \iid \ \Ncal(\muv_{S},\Sigma_S)$.}.

Depending on the structure of the covariance matrix $\Sigma_S$, the CCDP channel
 specializes in the following models:

\smallskip
\noindent
$\bullet$ {\bf The WRDP channel:} Corresponding to
\ea{
\Sigma_{S}=\Iv_M,
}
where $\Iv_M$ is the identity matrix of length $M$, that is, the channel states are independent white Gaussian sequences with zero mean and  unitary variance.

\smallskip
  \noindent
$\bullet$ {\bf The WSFD channel:}  Corresponding to
\ea{
\Sigma_{S}=\av^T \av,
}
for $\av=[a_1 \ldots a_m]$, that is,  each channel state sequence is equal to $S_m=a_m S^N$ where $S^N$ is a white Gaussian sequence with zero mean and unitary variance.
%
For this model, we further assume  $a_1 \leq  a_2 \leq \ldots \leq a_M$ without loss of generality.

\smallskip
\noindent
$\bullet$ {\bf The CCDP-ES channel:}
Corresponding to
\ea{
\Sigma_{S}=(1-\rho)\Iv_M + \rho  \onev_M^T \onev_M,
\label{eq:CCDP-ES}
}
where $\onev_M$ is the all-one row vector of size $M$,  that is, the channel states are Gaussian sequences
with zero mean, unitary variance and  pairwise correlation $\rho$
(the range of feasible values of $\rho$ is discussed later in Lem. \ref{lem:Feasible CCDP-ES}).

\medskip
In the following, we assume standard definitions of code, probability of error, achievable rate and capacity.
\begin{definition}{\bf Code and probability of error.}
A $(2^{NR},N)$ code for the CCDP channel  is defined by an encoding function $f(\cdot)$ with
\ea{
X^N=f(W,[S_1^N \ldots S_M^N]),
}
for $W\in \Wcal$, and $M$ decoding functions $g_m(\cdot)$ for
\ea{
\Wh_m=g_m(Y_m^N), \ m \in [1 \ldots M].
}
The probability of error  $P_e$ of a $(2^{NR},N)$ code for the CCDP channel is defined as
\ea{
P_e & = \max_{m \in [1 \ldots M]} \f 1 {2^{NR} } \sum_{w=1}^{2^{NR}} P_e(w)
\label{eq:def error probability} \\
P_e(w)& =\Pr \lsb g_m(Y_m^N) \neq  w| X^N=f(w,[S_1^N \ldots S_M^N]) \rsb. \nonumber
}
\end{definition}
Note that the error probability in \eqref{eq:def error probability} is  also averaged over all possible realizations of the state sequence vector $[S_1^N \ldots S_M^N]$.
%
\begin{definition}{\bf Achievable rate, capacity, and  approximate capacity.}
\label{def:approximate}
A rate $R$ is  said to be achievable on the CCDP channel if, for any $\ep>0$, there exists a code $(2^{NR'},N)$
such that $R'\geq R$ while $P_e \leq \ep$.
The capacity $\Ccal$ is defined as the supremum of all the achievable rates.
An inner bound $R^{\rm IN} \leq \Ccal$ and an outer bound $R^{\rm OUT} \geq \Ccal$ such that
\ea{
R^{\rm OUT}-R^{\rm IN} \leq \Delta,
}
for all channel parameters and for some constant $\Delta >0$ are said to determine the capacity to within an additive gap
of $\Delta$ bits--per--channel--use ($\bpcu$) or, for brevity, to characterize the approximate capacity to within $\Delta \ \bpcu$.
\end{definition}
In the following, we focus on determining the approximate capacity for the CCDP channel to within a small gap  for various parameter regimes.
Although partial, these results provide a tight characterization of capacity at high SNR.

\medskip

The channel model in \eqref{eq:compoundChannel} actually encompasses a larger class of compound channels with additive Gaussian states and additive Gaussian noise, as shown by the next lemma.
\begin{lem}{\bf Generalized channel model.}
\label{lem:Generalized Channel Model}
An $M$-receiver compound GP channel in which the  output at the $m^{\rm th}$ compound receiver is obtained as
\ea{
{Y'}_m^N = X'^N + S_m'^N + Z_m'^N, \quad m \in [1 \ldots M],
\label{eq:generalized channel}
}
where
\ean{
Z_m'^N & \sim \Ncal(\mu_Z, \sgs ) \\
[S_1'^N \ldots S_M'^N] & \sim  \iid \Ncal({\muv}_S,\Sigma_S),
}
while the input is subject to a power in \eqref{eq:power constraint}
can be reduced to the form of \eqref{eq:compoundChannel} without loss of generality.
\end{lem}

\begin{IEEEproof}
The  proof is provided in App. \ref{app:Generalized Channel Model}.
\end{IEEEproof}
In the following, we refer to the term $c$  in \eqref{eq:compoundChannel} as ``state gain'': although this term can be incorporated into the state covariance matrix $\Sigma_{S}$,
%
%
it is convenient to use
this  parameter to scale the variance of the state sequence across all outputs.

A simple but important observation is as follows.
\begin{lem}
\label{lem:capacity decreasing scaling fading}
The capacity of the CCDP channel is decreasing in the state gain $c$.
\end{lem}

\begin{IEEEproof}
The proof is provided in App.  \ref{app:capacity decreasing scaling fading}.
\end{IEEEproof}

The next lemma establishes the valid range of pairwise correlation for the  CCDP-ES channel.
\begin{lem}{\bf Feasible correlation for the CCDP-ES channel.}
\label{lem:Feasible CCDP-ES}
Let the matrix $\Sigma_S$ be defined as in \eqref{eq:CCDP-ES}:
%
then
$\Sigma_S$ is positive defined for $- 1/(M-1)  \leq \rho \leq 1$.
\end{lem}
\begin{IEEEproof}
See App. \ref{app:Feasible CCDP-ES}.
\end{IEEEproof}

\section{Related Results}
\label{sec:Related Results}

This section briefly reviews the results available in the literature which are relevant to the study of the CCDP channel.

\smallskip
\noindent
$\bullet$ {\bf  Gel'fand-Pinsker (GP) channel:}
%
%
The capacity for the GP channel \cite[Th. 1]{GelfandPinskerClassic} is obtained as
%
\ea{
\Ccal=\max_{P_{U,X|S}} \lb  I(Y; U) - I(U;S) \rb.
\label{eq:Capacity of GP channel}
}
The  expression in \eqref{eq:Capacity of GP channel} is convex in $P_{X|S,U}$ for a fixed $P_{U| S}$  which implies that $X$  can be chosen to be a deterministic function of $U$ and $S$.
On the other hand, this expression is neither convex nor concave in $P_{U|S}$ for a fixed $P_{X|S, U}$: for this reason, it is not easy to obtain a closed-form expression
of capacity or to evaluate it numerically.

\smallskip
\noindent
$\bullet$ {\bf  Writing on Dirty Paper (WDP) channel:}
%
One of the few channel models for which the maximization in \eqref{eq:Capacity of GP channel} is known in closed-form is the WDP channel \cite{costa1983writing}.
%
For this model the assignment
\ea{
& X \sim \Ncal(0,P),  \ X \perp S \nonumber\\
& U= X + \f {P} {P+1}S,
\label{eq:DPC assigment}
}
in \eqref{eq:Capacity of GP channel} recovers the point-to-point capacity.
This implies that full state pre-cancellation is possible regardless of the distribution of $S$.

\smallskip
\noindent
$\bullet$ {\bf  Carbon Copy onto Dirty Paper (CCDP)  channel:}
The CCDP channel is the compound extension of the WDP channel.
%
In \cite{LapidothCarbonCopying} the following bounds on the capacity of the $2$-receiver WRDP channel are shown.

\begin{thm}{\bf Inner and outer bounds for the $2$-receiver WRDP channel \cite[Th. 3, Th. 4]{LapidothCarbonCopying}. \\}
\label{th:Inner and outer bounds $2$-receiver CCDP with independent states }
Consider the $2$-receiver WRDP channel:
the capacity  of this model is upper bounded as
\ea{
R^{\rm OUT} = \lcb \p{
\f 14 \log\lb\f{1+P}{ c^2/4+1} \rb + \f 1 4 \lb \f{1+P+c^2+2 c \sqrt{P}}{c^2/4 +1}\rb  & c^2 < 4 \\
\f 14 \log(1+P)\\
\quad +\f 14 \log (1+P+c^2+2 c \sqrt{P} ) \\
\quad \quad  -\f 14 \log(c^2) & c^2 \geq 4,
} \rnone
\label{eq:outer lapidoth}
}
and lower  bounded as
\ea{
R^{\rm IN} = \lcb \p{
\f 12 \log \lb 1 + \f {P}{c^2/2+1}\rb   & c^2/2 \leq 1 \\
\f 12 \log \lb \f{P+c^2/2+1}{c^2}\rb  \\
\quad + \f 14 \log \lb \f {c^2} 2 \rb & 1 \leq c^2/2 <P+1 \\
\f 14 \log (P+1) & c^2/2 \geq P+1.
}
\rnone
\label{eq:inner lapidoth}
}
\end{thm}
The inner bound in \eqref{eq:inner lapidoth} is derived using a common codeword treating the channel state as noise and a private codeword for each user.
The private codewords employ lattice codes to pre-code the transmitted message against a linear combination of the two  state sequences.

The results in Th. \ref{th:Inner and outer bounds $2$-receiver CCDP with independent states } are also extended in \cite{LapidothCarbonCopying} to the case of any number of compound receivers.

\begin{thm}{\bf Outer bounds  for the $M$-receiver WRDP channel \cite[Eq. (31)]{LapidothCarbonCopying}. \\}
\label{th:Outer bounds  for the $M$-receiver WRDP channel}
Consider the $M$-receiver WRDP channel:
the capacity of this model is upper bounded as
\ea{
\Ccal
& \leq R^{\rm OUT} = \f 12 \log\lb P+c^2+2c \sqrt{P} \rb - \f{M-1}{2 M} \log c^2  \nonumber \\
&  \quad \quad -\f 1 {2 M}\log M - \lsb \f 1 {2M} \log \lb \f {c^2}{M(P+1)} \rb \rsb^+.
\label{eq:outer bound M independent states}
}
\end{thm}

\smallskip
\noindent
$\bullet$ {\bf  ``Writing on Fast Fading Dirt'' (WFFD)  channel:}
In the WSFD channel, the output at each receiver contains the same state sequence $S^N$  multiplied by a different scaling factor:
this models a WDP channel in which the channel state is affected by a slow fading process  known at the receiver.
The WFFD channel is the fast fading counterpart to the WSFD channel in which the channel output is obtained as
\ea{
Y^N=X^N+ c A^N \circ S^N + Z^N,
\label{eq:fast fading channel}
}
where $\circ$ indicates the Hadamard product, with
\ean{
S^N & \sim \iid  \Ncal(0,1) \\
A^N & \sim \iid P_A,
}
and where $A^N$ in known only at the receiver.
The terms $X^N$ and $Z^N$ in \eqref{eq:fast fading channel} are defined as in  \eqref{eq:compoundChannel}.
The capacity of the model in \eqref{eq:fast fading channel} is a special case of the capacity of the GP channel in \eqref{eq:Capacity of GP channel}.
In \cite{rini2014strongFading}, we derived alternative inner and outer bounds to the  expression in \eqref{eq:Capacity of GP channel} and show the approximate capacity for the case of antipodal fading realizations.

\begin{thm}{\bf Approximate capacity for the WFFD channel with Gaussian state and antipodal fading \cite{rini2014strongFading}.\\}
\label{th:Outer bound 2}
%
Consider the WFFD channel in \eqref{eq:fast fading channel} for the case in which $P_A$ is the uniform distribution over the set
$\{-1,+1\}$: the capacity for this model is upper bounded as
\ea{
\Ccal \leq R^{\rm OUT}  =
& \lcb \p{
 \f 1 2 \log(P+1)
 &  c^2 \leq 1 \\
 \f 12 \log(P+c^2+1) \\
\quad  - \f 14 \log(c^2+1)+\f3 2
& 1 < c^2 < P+1  \\
\f 1 4 \log(P+1)+2
& c^2 \geq P+1,
} \rnone
\label{eq:Outer bound 2 fading}
}
and the capacity lies to within a gap of $2 \ \bpcu$ from the outer bound in \eqref{eq:Outer bound 2 fading}.
\end{thm}

The outer bound in \eqref{eq:Outer bound 2 fading} can be approached by a transmission scheme in which the channel input is the superposition of two codewords: the base codeword treats the channel state as noise while the top codeword is pre-coded against $+S^N$.

\section{The Writing on Random Dirty Paper Channel}
\label{sec:The Writing on Random Dirty Paper (WRDP)}

In this section we derive the capacity of the $M$-receiver WRDP channel to within $2.25 \ \bpcu$:
we begin by considering the case of two compound receivers and successively extend this
result for any number of compound receivers.

\begin{figure}
\centering
\includegraphics[trim=1.5cm 0cm 1.5cm 0cm, width=0.45 \textwidth]{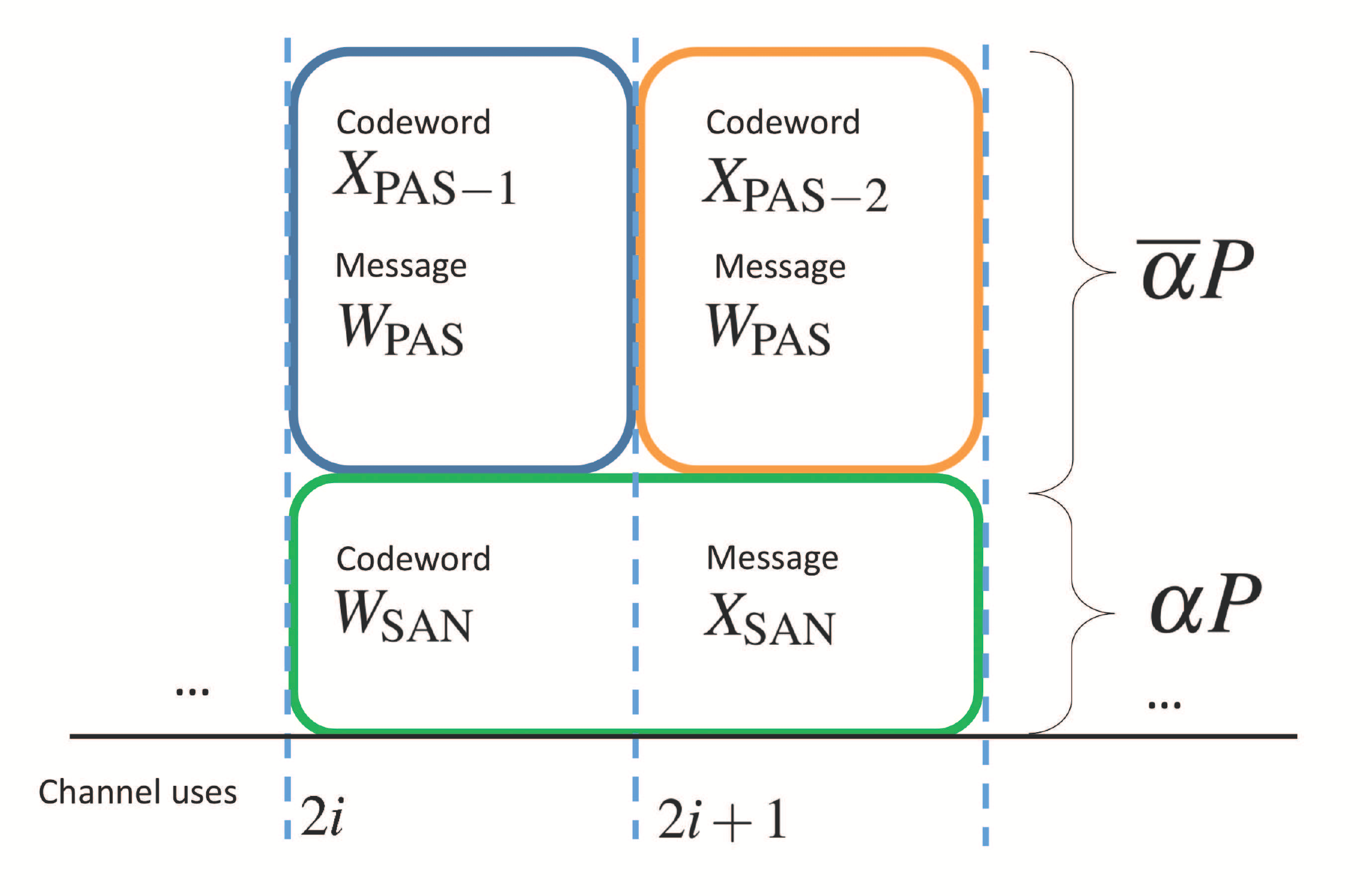}
\caption{A graphical representation of the capacity approaching  scheme in Th. \ref{th:Approximate capacity for the  $2$-receiver  CCDP with Gaussian independent states}.}
\label{fig:achievableSchemeSuperposition}
\end{figure}

\begin{thm}{\bf Approximate capacity for the  $2$-receiver  WRDP channel. \\}
\label{th:Approximate capacity for the  $2$-receiver  CCDP with Gaussian independent states}
Consider the $2$-receiver  WRDP channel: the capacity of this model is upper bounded as
\ea{
\Ccal \leq R^{\rm OUT}  =
& \lcb \p{
 \f 1 2 \log(P+1)
 &  c^2 \leq 1 \\
 \f 12 \log(P+c^2+1) \\
\quad  - \f 14 \log(c^2+1)+\f1 2
& 1 < c^2 < P+1  \\
\f 1 4 \log(P+1)+1
& c^2 \geq P+1,
} \rnone
\label{eq: outer bound CCDP bernoulli}
}
and the capacity  lies to within a gap of $1 \ \bpcu$ from the outer bound in \eqref{eq: outer bound CCDP bernoulli}.
\end{thm}
\begin{IEEEproof}%
%
When $c^2\leq 1$,  treating the channel states as additional noise attains the point-to-point capacity to within $1 \ \bpcu$.
When $P\leq1$, the point-to-point capacity is necessarily smaller than $1 \ \bpcu$ and thus the capacity of the WRDP channel is also smaller than $1 \ \bpcu$.
The proof for $c^2>1$  and $P>1$ is as follows.

\medskip
\noindent
\emph{ Converse:}
Using Fano's inequality and similarly to  \cite[Th. 3]{LapidothCarbonCopying}, we upper bound capacity as
\eas{
& N (R -\ep_N) \nonumber \\
& \leq \min_{m \in \{1,2\}}I(Y_m^N;W) \nonumber \\
& \leq \f 1 2 \lb H(Y_1^N)+H(Y_2^N) \rnone \label{eq:positive plus negative plus} \\
& \quad \lnone - H(Y_1^N|W)  - H(Y_2^N| W)\rb.
\label{eq:positive plus negative minus}
}
The positive entropy terms in \eqref{eq:positive plus negative plus} are  bounded as
\eas{
& H(Y_1^N)+H(Y_2^N) \nonumber \\
& \leq \f N 2 \log(P+c^2 + 2 c  \sqrt{P} +1) \nonumber  \\
& \quad + \f N 2 \log(P+ c^2+ 2 c  \sqrt{P} +1)+ N  \log 2 \pi e
\label{eq:sum entropy bound 1} \\
& \leq N  \log   (P+c^2 +1)+ N  \log 2 \pi e+1,
\label{eq:sum entropy bound 2}
}{\label{eq:sum entropy bound}}
where \eqref{eq:sum entropy bound 1} follows from the Gaussian Maximizes Entropy (GME) property and \eqref{eq:sum entropy bound 2} follows from the fact that
$2(P+c^2)\geq  (\sqrt{P}+c)^2$.
The negative  entropy terms in \eqref{eq:positive plus negative minus} are bounded as
\eas{
& -H(Y_1^N|W)  - H(Y_2^N| W) \nonumber \\
& \leq -H(Y_1^N,Y_2^N| W) \nonumber \\
& = -H(Y_2^N-Y_1^N,Y_2^N| W)
\label{eq:transform 1}\\
& = -H(c(S_2^N-S_1^N) +Z_2^N-Z_1^N, Y_2^N | W),
\label{eq:transform 11}
}{\label{eq:transform} }
where the change in variable in  \eqref{eq:transform 1}  has unitary Jacobian.
We continue the series of inequalities in \eqref{eq:transform} as
\eas{
%
\eqref{eq:transform 11} & = -H(c(S_2^N-S_1^N) +Z_2^N-Z_1^N| W)  \nonumber \\
& \quad \quad  - H(Y_2^N| S_2^N-S_1^N +Z_2^N-Z_1^N, W) \\
& \leq -H(c(S_2^N-S_1^N) +Z_2^N-Z_1^N) \nonumber \\
& \quad  \quad - H(Z_2^N| Z_2^N-Z_1^N) \\
& \leq  -\f N2 \log 2 \pi e ( 2 c^2 +2) - \f N2 \log 2 \pi e \f 12.
 \label{eq:bonuding difference states}
}
Combining \eqref{eq:sum entropy bound 2} and \eqref{eq:bonuding difference states} and for $c^2>1$ we have
\ea{
R^{\rm OUT} & =\f12 \log \lb P+c^2+1\rb \nonumber \\
& \quad \quad  -\f 14 \log \lb c^2 \rb +\f 12.
\label{eq:out independent 1}
}
The expression in \eqref{eq:out independent 1} is convex in $c^2$ with a minimum in ${c^2}^*=P+1$:
 following Lem. \ref{lem:capacity decreasing scaling fading}, decreasing the value of $c^2$ yields a channel with larger capacity.
For this reason, substituting $c^2$ in \eqref{eq:out independent 1} with  $\min\{c^2,P+1\}$ yields the  tighter outer bound.
This substitution produces the outer bound in \eqref{eq: outer bound CCDP bernoulli}.

\medskip
\noindent
\emph{Achievability:}
%
Consider the achievable strategy schematically presented in Fig. \ref{fig:achievableSchemeSuperposition}.
The channel input is obtained as the superposition of a bottom codeword
 and two top codewords.
The bottom codeword, $X_{\rm SAN}^N$ ($\rm SAN$ for \emph{State As Noise}) with power $\al P$,
carries the message $W_{\rm SAN}$ with rate $R_{\rm SAN}$.
This codeword treats the state sequences as additional noise and is decoded at both receivers.
The two top codewords, $X_{\rm PAS-1}^N$ and $X_{\rm PAS-2}^N$ ($\rm PAS$ for \emph{Pre-coded Against State}), both
have power $\alb P$ for $\alb=1-\al$  and  carry the message $W_{\rm PAS}$ at rate $R_{\rm PAS}$.
These two codewords are transmitted using time-sharing, each sent for half of the channel uses.
The codeword $X_{\rm PAS-1}^N$ is pre-coded against the state sequence $S_1^N$ as in the classical WDP channel and is decoded only at receiver~1.
Similarly, $X_{\rm PAS-2}^N$ is pre-coded against $S_2^N$ and decoded only at receiver~2.
Since the private codewords carry the same message, each compound receiver is able to decode both $W_{\rm SAN}$ and $W_{\rm PAS}$,
thus attaining the rate
\ea{
R^{\rm IN} = \f 12 \log \lb 1+ \f {\al P}{ c^2 +\alb P + 1}\rb  + \f 1 4 \log \lb 1+ \alb P \rb.
\label{eq: inner bound WRDP}
}
The expression in \eqref{eq: inner bound WRDP}  can be maximized over $\al$, the ratio between the power of the common and the private codewords.
When $P+1\geq c^2$,  the optimal value of $\alb$ yields $\alb P+1 = c^2 S_i,  \ i \in \{1,2\}$.
When $c^2 \geq P+1$, the optimal allocation yields $\al P=0$ and the transmission scheme reduces to pre-coding for each receiver for half of the time.
As a result of the optimization over $\al$ in \eqref{eq: inner bound WRDP}, we obtain the inner bound
\ea{
R^{\rm IN} = \lcb \p{
\f 12 \log \lb 1+ \f P {c^2 +1}\rb               & c^2 < 1 \\
\f 12 \log \lb 1+ c^2 + P\rb  \\
 \quad -\f 14 \log(c^2)- \f 12   & 1 \leq c^2 < P+1\\
\f 14 \log (P+1)               & c^2 \geq P+1.
}\rnone
\label{eq:inner bound 2user}
}
By comparing the expression in \eqref{eq: outer bound CCDP bernoulli} and \eqref{eq:inner bound 2user}, we conclude that the outer bound can be attained to within $1 \ \bpcu$.
\end{IEEEproof}

Next, we extend the result in Th. \ref{th:Approximate capacity for the  $2$-receiver  CCDP with Gaussian independent states} to the case of any number
of compound receivers.

\begin{thm}{\bf Approximate capacity for the  $M$-receiver WRDP channel. \\}
\label{th:Approximate capacity $M$-CCDP with Gaussian independent states}
Consider the $M$-receiver WRDP channel: the capacity for this model is upper bounded as
\ea{
& \Ccal \leq R^{\rm OUT} =
\label{eq: outer bound CCDP M} \\
& \lcb \p{
\f 12 \log \lb 1+\f  {P}{1+ c^2} \rb+2.25 &  M-1 > c^2 \\
 \f 1 {2M} \log(1+P)+ &  M-1 \leq c^2 < (M-1)(P+1) \\
 \quad +\f {M-1}{2 M } \log \lb {c^2} \rb+ 1.5  &  \\
 \f 1 {2M} \log(1+P)  +2   & c^2  \geq (M-1)(P+1),
} \rnone
\nonumber
}
and the capacity lies to within a gap of $2.25 \ \bpcu$ from the outer bond in \eqref{eq: outer bound CCDP M}.
\end{thm}
\begin{IEEEproof}
The converse proof is established using a recursion which extends on the outer bound derivation in the proof
of Th. \ref{th:Approximate capacity for the  $2$-receiver  CCDP with Gaussian independent states}.
The inner bound has the same spirt as the inner bound in Th. \ref{th:Approximate capacity for the  $2$-receiver  CCDP with Gaussian independent states}:
the channel input is obtained as the superposition of $M$ private codewords over a common codeword.
The common codeword treats the channel states as noise and is decoded at all receivers, while the private codewords are transmitted using
time-sharing.
Additionally, the $m^{\rm th}$ private codeword is pre-coded against the channel state at the $m^{\rm th}$ compound receiver and all convey the same message.
%
Similarly to \eqref{eq: inner bound WRDP}, the rate attainable with this strategy is
\ea{
R^{\rm IN} = \f 12 \log \lb 1+ \f {\al P}{ c^2 +\alb P + 1}\rb  + \f 1 {2M} \log \lb 1+ \alb P \rb,
\label{eq:attainable rate WRDP with M}
}
which can again be maximized over the power allocation parameter $\al$.
The full proof is provided in App. \ref{app:Approximate capacity $M$-CCDP with Gaussian independent states}.
\end{IEEEproof}
The result in Th. \ref{th:Approximate capacity $M$-CCDP with Gaussian independent states} essentially shows that it is not possible
to effectively pre-code against multiple independent channel state realizations.
Instead, a simple combination of time-sharing, superposition coding, and dirty paper coding is sufficient to closely approach capacity
and other, more complex, transmission  strategies such as joint binning, non-unique decoding and structure codes
provide no substantial improvements.

\begin{rem}{\bf Time-sharing VS code-sharing.}
\label{rem: ts}
The achievable strategy in the proof of Th. \ref{th:Approximate capacity $M$-CCDP with Gaussian independent states} can be improved upon by using code-sharing instead of time-sharing as follows:
\ea{
& X^N_{\rm SAN} \spc X_{\rm PAS-1}^N \nonumber \\ 
&  X^N_{\rm SAN} \spc X_{\rm PAS-2}^N \nonumber \\
& S_1^N \bin X_{\rm PAS-1}^N \nonumber \\
& S_2^N \bin X_{\rm PAS-2}^N   \nonumber  \\
& X_{\rm PAS-1}^N  \jbin X_{\rm PAS-2}^N,
\label{eq:all spc}
}
where $U^N \spc V^N$ indicates that $V^N$ is superimposed $U^N$  and $U^N \bin V^N$ indicates that $V^N$
is binned against $U^N$ as in \cite{rini2016unified}.
As in the Gaussian broadcast channel, in which superposition coding performs at most $1 \ \bpcu$ better
then time-sharing, the achievable strategy in \eqref{eq:all spc}
provides a bounded performance improvement  over the time-sharing  strategy used in the achievability proof of Th. \ref{th:Approximate capacity $M$-CCDP with Gaussian independent states}.
On the other hand, the simpler achievable strategy of Th. \ref{th:Approximate capacity $M$-CCDP with Gaussian independent states} can be more easily
optimized as a function of the channel parameters.
\end{rem}

\begin{rem}{\bf Non-unique decoding.}
\label{rem: non unique}
Indirect or non-unique decoding as in \cite{nair2010achievability} is not necessary for the result in Th.  \ref{th:Approximate capacity $M$-CCDP with Gaussian independent states}.
As argued in \cite{bidokhti2014non}, joint (unique) decoding   is sufficient to approach capacity to within a small gap.
It can  be shown that also for the scheme in Rem. \ref{rem: ts}, non-unique decoding does not provide rate improvements over unique decoding.
\end{rem}

\subsection{Discussion}
The relatively simple expression of the result in Th. \ref{th:Approximate capacity $M$-CCDP with Gaussian independent states}
is made possible by the assumption that the channel states all have equal variance.
When the states have the same variance, treating the channel state as noise attains the same rate at all compound receivers.
If the state sequences had different variance, we could improve upon the achievable scheme in Th. \ref{th:Approximate capacity $M$-CCDP with Gaussian independent states}
by employing partially common codewords, i.e. codewords which are decoded by a subset of receivers.
As an example consider the case of $M=3$ with channel states of increasing variance, i.e. $\var[S_1]<\var[S_2]<\var[S_3]$.
In this case, a codeword treating the channel state at user~2
as additional noise can also be decoded at receiver~1 but it cannot be decoded at receiver~3.
The use of partially common codeword  necessarily  introduces further constraints in the derivation of inner and outer bounds,
leading to a more complex expression of the approximate capacity.

\section{Writing on Slow Fading Dirt Channel}
\label{sec:Dirty Paper Channel  with Slow Fading Dirt}
This section investigates the capacity of the $M$-receiver WSFD channel: as in Sec. \ref{sec:The Writing on Random Dirty Paper (WRDP)},
we begin by considering the  case of two compound receivers and successively extend the analysis to the case of any $M$.
For the $2$-receiver  WSFD channel we  show the approximate capacity in all parameter regimes while, for the case of any number of
compound receivers, we are able to show capacity only under some additional conditions on the set of scaling coefficients $[a_1 \ldots a_M]$.
%
Since the WSFD channel models the WDP channel in which the channel state is multiplied by a slow fading process,
we refer to the term  $c a_m S^N$ as the \emph{fading-times-state term} at the $m^{\rm th}$
receiver\footnote{Note that this terminology is not coherent with the model definition in Sec. \ref{sec:Channel Model} but substantially facilitates the exposition of the results.}.
For the $2$-receiver  WSFD channel, we simplify the notation in \eqref{eq:compoundChannel} as
\ea{
Y_1^N & = X^N +  c S^N + Z_1^N \nonumber \\
Y_2^N & = X^N + c a S^N + Z_2^N,
\label{eq:2 WSFD model}
}
where $|a|\geq 1$ without loss of generality.
%
%
%
%
%
%
%

\begin{thm}{\bf  Approximate capacity for the $2$-receiver  WSFD channel.}
\label{th:Approximate capacity for the $2$-receiver  WSFD}

Consider the $2$-receiver  WSFD channel in \eqref{eq:2 WSFD model}: the capacity for this model is upper bounded as
{   \small
\ea{
& \Ccal \leq R^{\rm OUT}  = \label{eq: outer bound WFSFD a large} \\
& \lcb \p{
 \f 12 \log(P+1)
 & 1 \leq a <  1+\f 1 {\min \{\sqrt{P},c\}}  \\
 \f 14 \log(P+1)\\
 \ \ +\f 14 \log(\min \{P,c^2\}(a-1)^2+1 )
 & 1+\f 1 {\min \{\sqrt{P},c\}} \leq a \leq 2  \\
 \f 1 2 \log(P+2 c^2 (a-1)^2)^2\\
 \ \ -\f 1 4 \log(c^2 (a-1)^2+1)
& a\leq -1,a>2,  \ c^2a^2\leq P+1 \\
\f 1 4 \log(P+1)+\f 12
& a\leq -1,a>2, \ c^2a^2>P+1,
} \rnone
\nonumber
}}
and the capacity lies to within a gap of $4 \ \bpcu$ from the outer bound in \eqref{eq: outer bound WFSFD a large}.
\end{thm}

\begin{IEEEproof}
The proof requires a number of algebraic manipulations to simplify and compare inner and outer bound expressions:
  these details are omitted for brevity.

\medskip
\noindent
\emph{ Converse:}
With a derivation similar to the converse proof in Th. \ref{th:Approximate capacity for the  $2$-receiver  CCDP with Gaussian independent states},
we obtain the outer bound
\ea{
R^{\rm OUT}& = \f 14 \log(P+c^2+1)+\f 14 \log(P+ c^2 a^2+1) \nonumber \\
& \quad - \f 14 \log(c^2 (a-1)^2 + 1)+\f1 2.
\label{eq:m=2 passage 1}
}
The outer bound in \eqref{eq:m=2 passage 1} is close to capacity for $a\leq-1$ and $a>2$: in this regime, Lem. \ref{lem:capacity decreasing scaling fading} can be used to tightened the
expression in \eqref{eq:m=2 passage 1} by substituting $c^2$  with $\min\{c^2,c'^2\}$ in
\eqref{eq:m=2 passage 1} for
\ea{
c'^2 = \f {1+P} a^2 \leq c^2.
}
Further bounding of the expression \eqref{eq:m=2 passage 1} in the interval $a\in [1,2]$ yields
the expression in \eqref{eq: outer bound WFSFD a large}.

\medskip
\noindent
\emph{Achievability:}
For the model in \eqref{eq:2 WSFD model}, the achievable strategy employed
in Th. \ref{th:Approximate capacity for the  $2$-receiver  CCDP with Gaussian independent states}
can be  enhanced by pre-coding the common codeword  against the state sequence $S^N$ as in the GP channel.
Let $U_{\rm SAN}$ be the random variable corresponding to the binned codeword
and $X_{\rm SAN}$  the random variable associated with the transmitted codeword:
 this strategy attains the rate $R^{\rm SAN}$ for
\ea{
R^{\rm SAN}
& = H(U_{\rm SAN}|S) \label{eq:conditions common rate}\\
& \quad - \max \lcb H(U_{\rm SAN}|Y_1),H(U_{\rm SAN}|Y_2)\rcb.
 \nonumber 
}
For the expression in \eqref{eq:conditions common rate}, we consider the assignment
\ea{
& X_{\rm SAN}  \sim \Ncal (0,P), \ X_{\rm SAN} \perp S
\nonumber \\
& U_{\rm SAN}  =X_{\rm SAN} + k S,
\label{eq:assigment common rate}
}
with $k\in \Rbb$.
A partial optimization over $k$ in \eqref{eq:assigment common rate} yields the inner bound
\ea{
R^{\rm IN} & = \max_{\al \in [0,1]}  \f 12 \log \lb \f{\al P+1}  {  \f {\al P c^2 (1-a)^2}{P+c^2+1}+1+\alb P}  +1\rb \nonumber\\
& \quad \quad + \f 1 4 \log \lb \alb P\rb-1.
\label{eq:R CR}
}
As for the expression in \eqref{eq: inner bound WRDP}, the expression in \eqref{eq:R CR} can be optimized over $\al$, the power allocation parameter.

\medskip
\noindent
\emph{Gap to capacity:}
We separately consider three regimes of the fading coefficient $a$: a weak, medium and strong fading.

\smallskip
\noindent
$\bullet$ \underline{\emph{ Weak fading -- $a \in [1,1+1/\min\{\sqrt{P},c\})$:}}
Coding as in the WDP channel for the first compound receiver attains the rate
\ea{
 R^{\rm IN -WDP}
 & =\f 12 \log \lb P+1 \rb \nonumber \\
 & \quad \quad - \f 12 \log \lb  \f {P c^2 }{P+c^2+1}(1-a)^2+1 \rb,
 \label{eq:peer to peer inner bound}
}
at the second compound receiver.
In the given parameter regime, \eqref{eq:peer to peer inner bound} is to within  $1/2 \  \bpcu$ from the point-to-point capacity.

\smallskip
\noindent
$\bullet$ \underline{\emph{ Strong fading -- $a \in \Rbb \setminus [-1,2)$:}}
%
When $c^2a^2>P+1$, setting $\al=0$ in \eqref{eq:R CR} attains the outer bound in \eqref{eq: outer bound WFSFD a large} to within
$1/2 \ \bpcu$.
When $c^2a^2\leq P+1$, instead, the inner bound in \eqref{eq:R CR} for the assignment
\ea{
\al=\f {P+1-c^2 a^2} P,
}
is to within $3 \ \bpcu$ from the outer bound in \eqref{eq: outer bound WFSFD a large}.

\smallskip
\noindent
$\bullet$ \underline{\emph{ Medium fading -- $a \in [1+1/\min\{\sqrt{P},c\},2]$:}}
When either $P\leq 3$ or $c^2 \leq 3$, capacity can be attained to without $3 \ \bpcu$ by treating the channel states as noise.
For $P>3$ and $c^2>3$, consider the achievable scheme in \eqref{eq:R CR} for $\al=a-1$ which yields the inner bound
\ea{
R^{\rm IN} \geq  \f 1 4 \log (P)+\f 14 \log \lb   \f{P (a'P+1)}{(P(-a'^3+a'^2+a')+1)^2} \rb,
\label{eq:inner b}
}
where $a'=a-1$. The inner bound in \eqref{eq:inner b} is to within $3 \ \bpcu$ from the outer bound in \eqref{eq: outer bound WFSFD a large}.
\end{IEEEproof}
The result in Th. \ref{th:Approximate capacity for the $2$-receiver  WSFD} highlights the relationship between the WSFD channel, the WRDP channel and the WDP channel.
For small positive values of $a$, i.e. $1<a< 1+1/\min \{\sqrt{P},c\}$,  the WSFD channel behaves essentially
 as a WDP channel since binning as in the WDP channel performs close to the
AWGN capacity.
When $a\leq -1$ or $a>2$, instead, the WSFD channel behaves similarly to the WRDP channel and the coding strategy
 in Th. \ref{th:Approximate capacity for the  $2$-receiver  CCDP with Gaussian independent states} is sufficient to approach capacity.
This implies that the correlation between the channel states cannot be exploited to improve the communication rates in this regime.
For the remaining values of $a$, i.e. $1+1/\min \{\sqrt{P},c\} \leq a \leq 2$, the achievable scheme in \eqref{eq: outer bound WFSFD a large}
is necessary to approach capacity, as it allows for partial state pre-cancellation at both compound receivers simultaneously.

\begin{rem} As for Rem.  \ref{rem: ts}, in Th. \ref{th:Approximate capacity for the $2$-receiver  WSFD} a very simple transmission strategy is sufficient to closely approach capacity.
Although many coding techniques have been proposed for simultaneous state pre-cancellation, such as joint binning \cite{piantanida2010capacity} non-unique decoding \cite{nair2010achievability}, lattices codes \cite{LapidothCarbonCopying}, multiple description codes \cite{benammar2014multiple}, Th. \ref{th:Approximate capacity for the $2$-receiver  WSFD} shows that these strategies do not provide substantial improvements at high SNR.
\end{rem}

Let us return to the  strong fading conditions in Th. \ref{th:Approximate capacity for the $2$-receiver  WSFD}: when $c^2 a^2>P+1$,
capacity can be approached by transmitting toward each compound receiver as in the WDP channel for half of the time.
The next theorem extends this result to the case of any number of compound receivers.

\begin{thm}{\bf Outer bound and approximate capacity  for the ``strong fading'' regime and $a_1=0$.\\}
\label{th:Outer bound strong}
Consider the $M$-receiver WSFD channel with $a_1=0$,  $P\ \geq 1$ and
\eas{
& c^2 a_2^2 > P+1
\label{eq: strong fading 1} \\
& \f {a_m^2} {a_{m-1}^2} \geq P+1, \quad m \in [3 \ldots M],
\label{eq: strong fading 2}
}{\label{eq: strong fading}}
the capacity for this model is upper bounded as
\ea{
\Ccal \leq R^{\rm OUT} & = \f 1 {2M} \log(1+P) + \f 1 2 \log(M)+2,
\label{eq:outer bound strong}
}
and the capacity lies to within a gap of $\f 1 2 \log(M)+2 \ \bpcu$ from  the outer bound in \eqref{eq:outer bound strong}.
\end{thm}

\begin{IEEEproof}
The converse proof extends the outer bound in Th. \ref{th:Approximate capacity for the $2$-receiver  WSFD} in the strong fading regime
 to the case of any number of receivers $M$ by determining  conditions under which a recursion similar that  in Th. \ref{th:Approximate capacity $M$-CCDP with Gaussian independent states}
  can be established.
In the achievability proof,  the encoder transmits toward each compound receiver as in the WDP channel
for a portion $1/M$ of the time.
The full proof is provided in App. \ref{app:Outer bound strong}.
\end{IEEEproof}

The strong fading conditions in Th. \ref{th:Outer bound strong}  are intuitively understood  through the deterministic binary linear approximation of  \cite{bresler_tse} of a Gaussian network: this model is particularly useful in understanding the interaction between  the different signals producing a channel output through a powerful visualization.
We briefly introduce this model here, solely for illustrative purposes: more details can be found in  \cite{bresler_tse} and in the
 related literature.
Consider  the  binary vector channel
\ea{
\Yo_{m}^N={\Sv}^{k-n_p} \Xo_{k}^N + \Sv^{k-n_{a_m}} \So_k^N,
\label{eq:deterministic}
}
where $\Sv$ is a binary matrix with $\Sv_{ij} =\delta_{i-1,j}$ for $(i,j) \in [1 \ldots m]^2$ and $\Xo_k^N$ and $\So_k^N$ are the first $k$ bits of the binary expansion of $X^N$ and $S^N$
respectively.
Also let $n_p=\lceil \log(P) \rceil$ and $n_{a_m}=\lceil \log(c a_m) \rceil$ and $k=\max\{n_p,n_{a_m}\}$ so that the  multiplication by ${\Sv}^{k-n_p}$ erases all but the $n_p$ most
significant bits of the binary vector $\Xo_k, \ k \in [1 \ldots N]$.
Similarly, the multiplication $\Sv^{k-n_{a_m}} \So_k$ erases all but the $n_{a_m}$ most significant bits of $\So_k$.

\begin{figure}
\centering
\includegraphics[trim=2.75cm 0cm 0cm 0cm, width=.6    \textwidth]{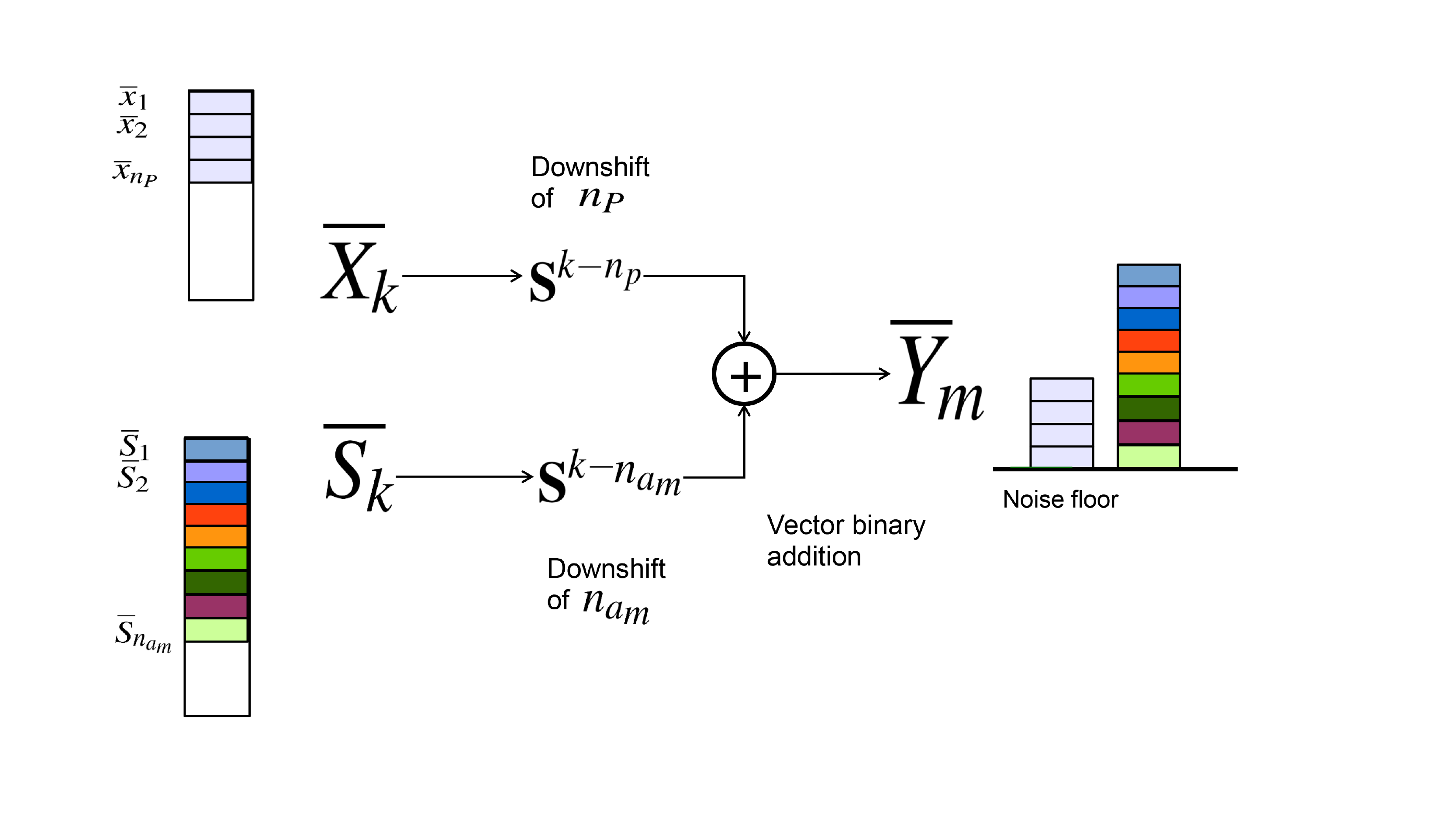}
\vspace{-1 cm}
\caption{The linear deterministic approximation of the WSFD channel.}
\label{fig:deterministic}
\end{figure}

The model in \eqref{eq:deterministic} is also represented Fig. \ref{fig:deterministic}: from a high-level perspective, it approximates the
behaviour of its Gaussian counterpart with a
 binary deterministic channel by replacing the additive noise with erasures and approximating the sum over $\Rbb$
 with the XORing of binary vectors.

\medskip

\begin{figure}
\centering
\includegraphics[trim=1.5cm 0cm 0cm 0cm, width=.57 \textwidth]{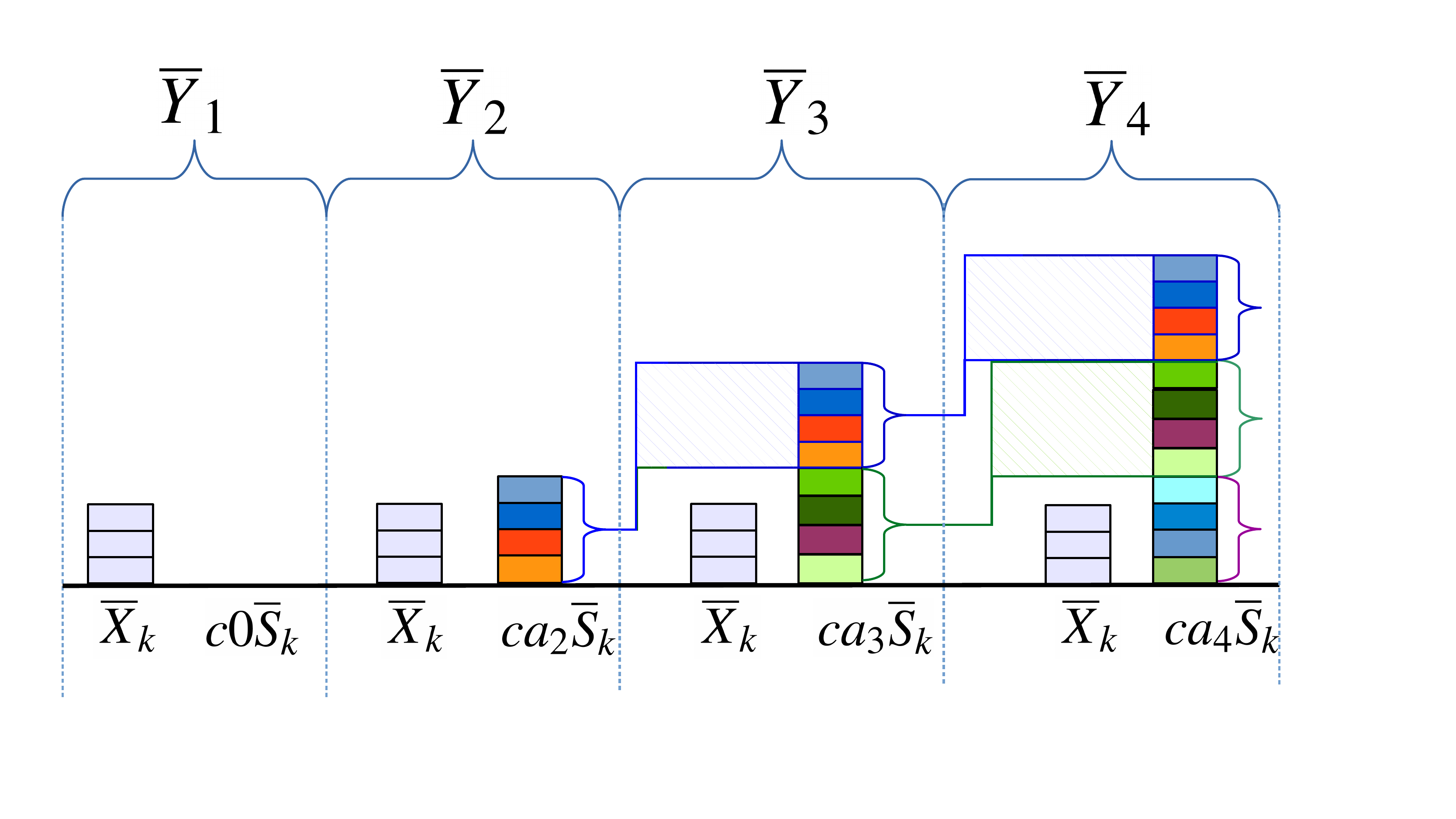}
\vspace{-1 cm}
\caption{The linear deterministic approximation of the 4-receiver WSFD channel in the ``strong fading'' regime.}
\label{fig:deterministic3}
\end{figure}

Through the approximation in Fig. \ref{fig:deterministic} we can better visualize the strong fading conditions
in Th. \ref{th:Outer bound strong}.
Consider Fig. \ref{fig:deterministic3} which represents, in vertical sections, the output at each compound receiver in the linear deterministic
approximation of the $4$-receiver  WSFD.
Each output  is obtained from a  different set of bits in the vector $\So_k$:  as $m$ increases, the value of $a_m$  increases and
more bits of  $\So_k$ appear above the noise floor.
When two coefficients $a_m$ and $a_{m+1}$  are sufficiently close,  the channel input sums with similar portions of the vector $\So_k$
and the encoder is potentially able to pre-code its transmitted codeword for these two receivers simultaneously.
When $a_m$ and $a_{m+1}$ are sufficiently different, instead, the channel input sums with two independent portions of the sequence $S_k$ and the channel substantially reduces to a WRDP channel.
This occurs when the ratio of $a_{m+1}$ and $a_m$ is larger than the magnitude
of the channel input, as illustrated in Fig. \ref{fig:deterministic3}, which approximatively corresponds to the conditions in \eqref{eq: strong fading}.

The condition $a_1=0$  in Th. \ref{th:Outer bound strong} is imposed only  in order to obtain a relatively
intuitive expression for the ``strong fading'' regime   as in Fig. \ref{fig:deterministic3}.
The next theorem presents a more general version of Th. \ref{th:Outer bound strong}.

\begin{lem}{\bf Outer bound and approximate capacity  for the  ``strong fading'' regime.}
\label{lem:Outer bound strong v2}

Consider a $M$-receiver WSFD channel and let $\Delta_m=a_m-a_1$ for $m \in [2\ldots M]$: if
\eas{
& c^2 \Delta_2^2> \max\{P+1,a_2^2\}\\
& c^2 \Delta_i^2 >1, \quad i>2 \\
& \sum_{i=2}^{m-1} \Delta_i^2 \geq  \gamma a_m^2, \quad \ \ \ m\in [2\ldots M]   \\
& \Delta_m^2 \geq \gamma  P \sum_{i=2}^{m-1} \Delta_i^2, \quad  m\in [3\ldots M],
}{\label{eq: strong fading v2}}
for some $\gamma>0$, then the capacity is upper bounded as
\ea{
\Ccal \leq R^{\rm OUT} & = \f 1 {2M} \log(1+P) + \f 1 2 \log(M)+\f 1 2 \log(\gamma)+2,
\label{eq:outer bound strong v2}
}
and the capacity lies to within a gap of $\f 1 2 \log(M)+\f 1 2 \log(\gamma) +2 \ \bpcu$ from the outer bound in \eqref{eq:outer bound strong v2}.
%
%
\end{lem}

\begin{IEEEproof}
The complete proof is provided in App. \ref{app:Outer bound strong v2}.
\end{IEEEproof}

It can be verified that the conditions in \eqref{eq: strong fading v2} reduce to the conditions in  \eqref{eq: strong fading} when letting $a_1=0$.
\subsection{Discussion}
In \cite{rini2014strongFading} we  determine the approximate capacity of the WFFD channel with antipodal fading: it is interesting to compare the different effects of slow and fast fading on the capacity of the WDP channel for the antipodal fading distribution.
By letting  $a=-1$ in Th. \ref{th:Approximate capacity for the $2$-receiver  WSFD} and comparing the resulting expression with
\eqref{eq:Outer bound 2 fading} in Th. \ref{th:Outer bound 2}, we see that the two regions are substantially identical.
This equivalence is rather interesting as one would not expect fast and slow fading to have roughly the same effect on the capacity of the WDP channel.
In the WFFD channel, from a high-level perspective, each typical realization of the fading distribution can be thought of as corresponding to a compound receiver.
Accordingly, the number of compound receivers in the WFFD channel can be imagined as growing exponentially with the blocklengh.
In the WSFD channel, instead, the number of compound receivers is fixed.

In the capacity approaching inner bound in Th. \ref{th:Approximate capacity for the $2$-receiver  WSFD}, the transmitter
pre-codes against the sequence $+S^N$ half of the time and against the sequence $-S^N$ for the other half of the time.
On the other hand, in the capacity approaching inner bound in Th. \ref{th:Outer bound 2}, the transmitter pre-codes against the
 realization $+S^N$ and each compound receiver observes $+S^N$ half of the time and  $-S^N$ the other half of the time on average.
In this sense, then, both capacity approaching schemes for the WSFD and WFFD channel serve half of the compound receivers
 at each time instance on average, so that the two schemes attain the same overall performance.
\section{Carbon Copy onto Dirty Paper Channel with Equivalent States}
\label{sec:The general  Carbon Copy onto Dirty Paper (CCDP) channel}

In this section we derive the approximate capacity for the $M$-receiver CCDP-ES:
as for the previous sections, we begin by studying the case of two compound receivers and successively investigate the general case.

Consider   $2$-receiver  CCDP channel in  \eqref{eq:compoundChannel}
and let the state covariance matrix $\Sigma_S$ be parameterized as
\ea{
\Sigma_S = \lsb  \p{
1               & \rho \sqrt{Q} \\
\rho \sqrt{Q}   & Q
}\rsb,
\label{eq:2CCDP sigma}
}
for some $Q\geq 1$ and $-1\leq \rho \leq 1$: the channel input/output relationship can be rewritten as
\eas{
Y_1^N & =  X^N + c S_1^N +Z_1^N  \nonumber \\
      & = X + c \lb \ka  S_c^N + \sqrt{1-\ka} \St_1^N \rb + Z_1^N \\
Y_2^N & =  X^N + c S_2^N +Z_2^N  \nonumber \\
      & = X + c \sqrt{Q} \lb \f{\rho} {\ka} S_c^N  + \sqrt{1-\f {\rho^2} {\ka^2} } \St_2^N \rb + Z_2^N,
}{\label{eq:common noise}}
for some $S_c^N,\St_1^N,\St_2^N \sim  \iid   \Ncal(0,1)$ and any $\ka \in [\rho,1]$.

%
The expression in \eqref{eq:common noise} shows how the CCDP channel can be treated as a combination of WRDP and WSFD channels: part of the state, $S_c^N$ is a common state while part of the state is independent from the state of the other user $\St_1^N$ and $\St_2^N$  respectively.

\begin{thm}{\bf Approximate capacity for a class of  $2$-receiver  CCDP-ES channel.}
\label{th:Approximate capacity for the  $2$-receiver  CCDP correlated}
Consider the $2$-receiver  CCDP-ES channel: the capacity for this model can be upper bounded as
\ea{
& \Ccal \leq  R^{\rm OUT}  =
 \lcb \p{
 \f 1 2 \log(P+1)
 &  c^2 \rhob^+ \leq 1 \\
 \f 12 \log(P+c^2+1) & 1 < c^2 \rhob^+ < P+1   \\
\quad  - \f 14 \log(c^2)+\f1 2
\\
\f 1 4 \log(P+1)+\f 12
& c^2\rhob^+ \geq P+1,
} \rnone
\label{eq: outer bound $2$-receiver  CCDP correlated}
}
for $\rhob^+=1-\max\{0,\rho\}$ and the capacity is to within $1 \ \bpcu$ from  the outer bound in \eqref{eq: outer bound $2$-receiver  CCDP correlated}.
\end{thm}
\begin{IEEEproof}
From  \eqref{eq:common noise}, we see that the CCDP-ES channel output can be rewritten as
\ea{
Y_m^N & = X^N + c \lb \sqrt{\rho} \ S_c^N + \sqrt{1-\rho} \ \St_m^N \rb + Z_m^N,
\label{eq:independent additive noise correlated}
}
with  $m \in \{1,2\}$ by letting $Q=1$ and fixing $k =\sqrt{\rho}$ in \eqref{eq:common noise}.

The achievability in Th. \ref{th:Approximate capacity for the  $2$-receiver  CCDP correlated}  follows the achievability
in Th. \ref{th:Approximate capacity for the  $2$-receiver  CCDP with Gaussian independent states} by additionally pre-coding the codeword $X_{\rm SAN}^N$ against the common state sequence $S_c^N$ in \eqref{eq:independent additive noise correlated}.
The converse is similarly  obtained from the converse of Th. \ref{th:Approximate capacity for the  $2$-receiver  CCDP correlated}  by additionally providing the common state sequence $S_c^N$ as a genie-aided side information to all the receivers.
The complete proof is provided in App. \ref{app:Approximate capacity for the  $2$-receiver  CCDP correlated}.
\end{IEEEproof}
Note that the result in Th. \ref{th:Approximate capacity for the  $2$-receiver  CCDP correlated}  coincides with the results in Th. \ref{th:Approximate capacity for the  $2$-receiver  CCDP with Gaussian independent states} when $\rho$ is negative.
This shows that the capacity of the channel with  negative correlation is substantially the same as the capacity of the channel with independent channel states.

The result in Th. \ref{th:Approximate capacity for the  $2$-receiver  CCDP correlated} can be  extended to the case of any number of receivers $M$ when the channel states have the same variance and the same pairwise correlation.
\begin{thm}{\bf Approximate capacity for a class of $M$-receiver CCDP-ES channel.}
\label{th:Approximate capacity for the  $M$-CCDP with Gaussian dependent states}
Consider the  $M$-receiver CCDP-ES: then capacity of this model is upper bounded as
\ea{
& \Ccal \leq  R^{\rm OUT}  = \\
& \lcb \p{
\f 12 \log \lb 1+\f  {P}{1+ c^2} \rb+\f 9 4    &  M-1 \geq  c^2 \rhob^+ \\
 \f 1 {2M} \log(1+P) & {\small M-1 < c^2 \rhob^+  \leq (M-1)(P+1) }\\
 \quad +\f {M-1}{2 M } \log \lb {c^2} \rb+ \f 3 2  &  \\
 \f 1 {2M} \log(1+P)  +2   & c^2 \rhob^+  > (M-1)(P+1),
} \rnone
\label{eq:outer M CCDP rho}
}
for $\rhob^+=1-\max\{0,\rho\}$ and the capacity is to within $2.25 \ \bpcu$ from  the outer bound in \eqref{eq:outer M CCDP rho}.
\end{thm}
\begin{IEEEproof}
As in \eqref{eq:common noise} and for $\rho>0$, each channel output can be rewritten as
\ea{
Y_m^N & = X^N + c (S_c^N + \St_m^N) + Z_m^N,
\label{eq:factor two}
}
for  $S_c,\St_m \sim  \Ncal(0,1), \ m \in [1 \ldots M]$.
The capacity  result in Th. \ref{th:Approximate capacity for the  $M$-CCDP with Gaussian dependent states} is  obtained by
 adapting the derivation in Th. \ref{th:Approximate capacity $M$-CCDP with Gaussian independent states} as follows:
%
for the achievability part, the common codeword is pre-coded against the common component of the state sequence $S_c^N$.
%
In the converse, $S_c^N$ is provided as genie-aided side information to all the receivers.
\end{IEEEproof}
When $\rho<0$, the channel output of the CCDP-ES can be equivalently expressed  as
\ea{
& Y_m^N  = X^N + Z_m^N+  c \lb - \sqrt{|\rho|} \sum_{j=1}^{m-1} \St_{jm}^N \rnone
\label{eq:negatie}\\
& \quad \quad \lnone +\sqrt{|\rho|} \sum_{j=m+1}^{M} \St_{mj}^N
+ \sqrt{1- (M-1)|\rho|} \St_{mm}^N \rb, \nonumber
}
for $\St_{i,q}, \sim \Ncal(0,1), \ i,q \in [1 \ldots M]^2$.
Note that each term $\St_{mj}$ appears with a negative sign in the expression of $Y_m$ and with  a negative sign in the expression of $Y_j$, thus yielding a negative correlation among each two state terms $S_m$ and $S_j$.
The expression in \eqref{eq:negatie} intuitively shows why no common channel state term emerges from negatively correlated channel states.
Note that the decomposition in \eqref{eq:negatie} also ostensibly motivates why the minimum negative correlation  $\rho$ is $-1/(M-1)$ as in Lem \ref{lem:Feasible CCDP-ES}, since $\St_m$ must contain $M-1$ terms to be negatively correlated with all the remaining  channel states.


We conclude by showing the approximate capacity of the $2$-receiver CCDP channel with independent states with unequal variance, obtained by setting $\rho=0$ in \eqref{eq:2CCDP sigma}.

\begin{thm}{\bf $2$-receiver  CCDP channel with independent states with unequal variance.}

Consider $2$-receiver  CCDP, the capacity for this model can be upper bounded as
\ea{
& \Ccal \leq R^{\rm OUT}  =
\label{eq: outer bound CCDP bernoulli general} \\
& \lcb \p{
 \f 1 2 \log(P+1)
 &  c^2  \sqrt{Q} \leq 1 \\
\f 1 4 \log(1+P+c^2)\\
 \quad +\f 14 \log(1+P+c^2Q) & 1 < c^2 \sqrt{Q} < P+1  \\
\quad \quad  - \f 14 \log(c^2(1+Q)+1)+\f3 2
 \\
\f 1 4 \log(P+1)+2
& c^2 \sqrt{Q} \geq P+1,
} \rnone
\nonumber
}
and the capacity $\Ccal$ is to within a gap of $2 \ \bpcu$ from $R^{\rm OUT}$ in \eqref{eq: outer bound CCDP bernoulli general} for $c^2 \sqrt{Q} \geq P+1$.
\end{thm}
\begin{IEEEproof}
The proof follows the same lines as the proof of Th. \ref{th:Approximate capacity for the  $2$-receiver  CCDP with Gaussian independent states}.
\end{IEEEproof}

\section{Conclusions}
\label{sec:Conclusion}

In this paper we investigate the capacity of the ``Carbon Copying onto Dirty Paper'' (CCDP) channel,  the compound version of the classic ``Writing on Dirty Paper'' (WDP) channel in which the channel output at each compound receiver is obtained as the sum of the input, Gaussian noise and one of $M$ possible channel Gaussian state sequences.
The state sequences are  anti-causally known the transmitter but unknown at the receivers.
For this model, we focus  on two scenarios: the case \emph{i.i.d.}  state sequences and the case in which the state sequences are scaled versions of a given sequence.

The case of \emph{i.i.d.} state sequences arises from the WDP channel in which multiple interferers have the potential of affecting the transmission but the transmitter has no knowledge of which one  eventually appears in the channel output.
The case of states being different scaling of the same sequence models the WDP channel in which the state sequence is multiplied by a slow fading coefficient which is known at the receiver but unknown at the transmitter.

For the case of \emph{i.i.d.} state sequences,  we derive capacity to within a constant gap for any number of compound receivers and any channel parameter.
In particular, we show that capacity can be approached with a rather simple strategy in which the input is composed of the superposition of two codewords:
a bottom codeword treating the channel states as noise and the top codeword  pre-coded against the channel state experienced at each compound receiver for a portion $1/M$ of the time.

For the case in which  the state sequences are scaled version of the same sequence, we derive the capacity to within a constant gap  for the case of two compound receivers  and  extend this result to the case of any number of receivers under some conditions on the scaling coefficients which we denote as ``strong fading'' regime.
In this parameter regime, the scaling coefficients are exponentially spaced apart and the encoder is unable to simultaneously pre-code against multiple scaling coefficients.
%
The capacity of the CCDP channel in which the state have any jointly Gaussian distribution remains an interesting open problem.

%
\bibliographystyle{IEEEtran}
\bibliography{steBib1,steBib2}
%
\appendices
\section{Proof of Lem. \ref{lem:Generalized Channel Model}}
\label{app:Generalized Channel Model}
The mean of the noise and the channel states can be removed from the channel outputs and
 each output can be scaled so that the noise variance becomes unitary, i.e. \ea{
Y^N_m=\f 1 {\sigma} \lb {Y'_m}^N -\mu_S-\mu_m \rb.
\label{eq:output scaling}
}
Since the transformation in \eqref{eq:output scaling} is a one-to-one transformation, it does not affect capacity.
The scaling of the channel input in  \eqref{eq:output scaling} can be incorporated into the  power constraint in \eqref{eq:power constraint} by defining
\ea{
X^N = \f 1 {\sigma} {X'}^N,
}
and letting $P=P'/\sgs$.
Similarly, the  parameter $c$ and the $S_m^N$ can be defined as
\eas{
c     & =\sqrt{ \f{\var[S_{\min}']}{\sgs}} \\
S_m^N & =  \f { {S'}_m^N }{\sqrt{\var[S_{\min}']}},
}
where $S_{\min}'$ is the state with the smallest variance, to match the CCDP channel definition in \eqref{eq:compoundChannel}.
Finally, since the state distribution is symmetric around the mean, we can take  $c$ to be positive  without loss of generality.

\section{Proof of Lem. \ref{lem:Feasible CCDP-ES}}
\label{app:Feasible CCDP-ES}

For the matrix in \eqref{eq:CCDP-ES},
the leading principal minor of order $m$ can be obtained through the matrix determinant lemma as
\ea{
\det((1-\rho)\Iv_{m}+\rho \ones_{m}^T \ones_{m} )=(1-\rho)^m \lb 1+ \f{m \rho} {1-\rho}\rb,
}
which is non-negative defined for
\ea{
-\f 1 {m-1} \leq \rho \leq 1.
}
Consequently, all the leading principal minors of the matrix in \eqref{eq:CCDP-ES}  are all positive when
\ea{
\min_{m} \lcb -\f 1 {m-1} \rcb =-  \f 1 {M-1} \leq \rho \leq 1.
\label{eq:princ minors}
}

\section{Proof of Lem. \ref{lem:capacity decreasing scaling fading}.}
\label{app:capacity decreasing scaling fading}
The state sequence vector $\So^N=[S_1^N \ldots S_M^N]$ can be expressed as
\ea{
\So^N=\So_{1}^N+\So_2^N,
}
where
\eas{
\So_1^N & \sim \iid \Ncal(0, \ga \Sigma_S)\\
\So_2^N & \sim \iid \Ncal(0, \gao \Sigma_S),
}
for $\So_1^N \perp \So_2^N$ and  $\gao=1-\ga$.
Consider now the CCDP channel in which $\So_2^N$ is provided as a genie-aided side information to the transmitter and all the compound receivers: the capacity of this channel is necessarily larger than the capacity of the original channel since this extra knowledge can be ignored.
The $m^{\rm th}$ compound receiver in the enhanced channel can produce the equivalent channel output
\ea{
\Yt_{m}^N
& =Y_{m}^N-c S_{2,m}^N \nonumber \\
& = X^N + c S_{1,m}^N + Z_m^N.
\label{eq:equivalent channel out}
}
For the CCDP channel with outputs as in \eqref{eq:equivalent channel out},  $\So_2^N$  acts as a common information, independent from the all other random variables, and thus the knowledge of $\So_2^N$ at all terminals does not increase capacity.
From Lem. \ref{lem:Generalized Channel Model}, we have that the CCDP channel with channel outputs as in \eqref{eq:equivalent channel out}
is statistically equivalent to the CCDP channel with state gain
\ea{
\ct=c \sqrt{\ga} \leq c,
\label{eq:ct}
}
and covariance matrix $\Sigma_S$.
We thus conclude that the capacity of the CCDP channel with state gain $c$ and common information $\So_2^N$ is equivalent to the
capacity of the CCDP channel with state gain is $\ct$ in \eqref{eq:ct}.
Accordingly, capacity is decreasing in $c$.
\section{Proof of Th. \ref{th:Approximate capacity $M$-CCDP with Gaussian independent states}.}
\label{app:Approximate capacity $M$-CCDP with Gaussian independent states}

As for the proof of Th. \ref{th:Approximate capacity for the  $2$-receiver  CCDP with Gaussian independent states}, when $P \leq 3$ and $c^2 \leq 3$ capacity can be attained to within $2 \ \bpcu$.
%
For $P>3$ and $c^2>3$, achievability  and converse proofs are derived as follows.

\noindent
\emph{Converse:}
As in \cite[App. 3.C]{LapidothCarbonCopying}, we write
\eas{
N(R-\ep)
& \leq \min_{m \in [1 \ldots M]}  I(Y_m^N;W)\\
& \leq \f 1 M \sum_{m=1}^M I(Y_m^N;W) \\
& \leq  \max_{m \in [1 \ldots M]} H(Y_m^N) - \f 1 M \sum_{m=1}^M  H(Y_m^N|W)  \\
& \leq  \f N 2 \log (P+c^2+2 c \sqrt{P}+1) \label{eq:label M case} \\
& \quad \quad +\f N 2 \log (2 \pi e) - \f 1 M  \sum_{m=1}^M  H(Y_m^N|W).
 \nonumber
}
The negative entropy terms in \eqref{eq:label M case} are bounded through recursion: we begin by defining two terms involved in the recursion:
\eas{
T_m & = \sum_{i=m}^M  H(Y_i^N|W)
\label{eq:def Tm}\\
\Delta^N_m &  = S_{m}^N-S_{m+1}^N, \ m\in [1\ldots M].
\label{eq:delta def}
}{\label{eq:def wrdp}}
In order to simplify the derivation, we also set the noise terms to be identical, i.e. \ea{
Z_1^N=Z_2^N=\ldots=Z_M^N.
\label{eq:ass equal noise}
}
%
Eq. \eqref{eq:ass equal noise} follows because the compound receivers are not allowed to cooperate and thus the joint distribution of
the noise terms $\{ Z_m,  \ m \in [1 \ldots  M]\}$ does not affect capacity.

With a derivation similar to that in \eqref{eq:sum entropy bound} and  using the assumption in \eqref{eq:ass equal noise}, we write
\eas{
- T_1&  = -H(Y_1^N|W)-H(Y_2^N|W)-T_3\\
    & \leq  -H(c(S_1^N-S_2^N) +Z_2^N-Z_1^N, Y_2^N | W)-T_3 \\
    & =  -H(c \De_1)-H(Y_2^N | \De_1,W)-T_3.
}{\label{eq:recursion step 1}}
The passage in \eqref{eq:recursion step 1} can be recursively repeated $M$ times where, at  $K^{\rm th}$ recursion step with $K \in [2, \ldots M]$, we have
\ean{
-T_1 & \leq   -\sum_{m=1}^{K} H(c \Delta^N_m| \Delta^N_1 \ldots \Delta^N_{m-1}) \nonumber \\
& \quad \quad \quad   - H(Y_{K}^N|\Delta^N_1 \ldots \Delta^N_{K-1},W) - T_{K+1} \\
& =  -\sum_{m=1}^{K} H(c \Delta^N_m| \Delta^N_1 \ldots \Delta^N_{m-1})  \nonumber \\
& \quad \quad \quad   - H(Y_{K}|\Delta^N_1 \ldots \Delta^N_{K-1},W) \\
& \quad \quad \quad \quad \quad \quad      - H(Y_{K+1}|W) - T_{K+2} \\
& \leq  -\sum_{m=1}^{K} H(c \Delta^N_m| \Delta^N_1 \ldots \Delta^N_{m-1}) \\
& \quad \quad \quad   - H(Y_{K}, Y_{K+1}|\Delta^N_1 \ldots \Delta^N_{K-1},W) - T_{K+2}\\
& =  -\sum_{m=1}^{K+1} H(c \Delta^N_m| \Delta^N_1 \ldots \Delta^N_{m-1}) \\
& \quad \quad \quad   - H(Y_{K+1}|\Delta^N_1 \ldots \Delta^N_{K},W) - T_{K+2}.
}
 By proceeding in this manner up to $K=M$, we come to the bound
\ea{
- T_1
& \leq \sum_{m=2}^{M} -H(c \Delta^N_m| \Delta^N_1 \ldots \Delta^N_{m-1}) \nonumber \\
& \quad \quad \quad    - H(Y_M^N|\Delta^N_1 \ldots \Delta^N_{M},W) \nonumber \\
& \leq \sum_{m=2}^{M} -H(c \Delta^N_m| \Delta^N_1 \ldots \Delta^N_{m-1}) - H(Z_M^N) \nonumber \\
& \leq \sum_{m=2}^{M} -H(c \Delta^N_m| \Delta^N_1 \ldots \Delta^N_{m-1}) - \f N 2 \log(2 \pi e).
\label{eq:last sum}
}
We next evaluate the different terms in the summation  \eqref{eq:last sum}, i.e.
\ean{
& H(c \Delta^N_m| \Delta^N_1 \ldots \Delta^N_{m-1})=  \\
& \quad \quad \f 1 2 \log(c^2) + H(\Delta^N_m| \Delta^N_1 \ldots \Delta^N_{m-1}),
}
the correlation matrix of the vector $\Delta^N=[\Delta^N_1 \ldots \Delta^N_M]$ is
\ean{
\Sigma_{\Delta^N} =  \lsb\p{
2       & -1        &  0 & 0    &  &  \ldots & 0 \\
-1      &  2        & -1 &     & & \ldots & 0 \\
0       & -1        &  2 & - 1 & & \ldots & 0 \\
\vdots  &           & \ddots   & \ddots  & \ddots  & & \vdots  \\
 &&0 & -1 &2 & -1 & 0\\
\vdots   &&&0 & -1 &2 & -1\\
0       & \ldots  &&&0 & -1 &2 \\
}
\rsb,
}
and thus we conclude that
\ea{
 & - H( \Delta^N_m| \Delta^N_1 \ldots \Delta^N_{m-1}) \nonumber \\
 & =  -\f 12 \log \lb 2 -   \lsb - 1 \ldots -1 \rsb   \cdot \lsb\p{2 & -1 &  0\\ -1 & \ddots & \ddots \\ 0 &   \ddots  & } \rsb \cdot \lsb \p{ -1 \\ \vdots \\ -1 \\ } \rsb  \rb  \nonumber \\
& =- \f 12 \log \lb 2  -   \f {i-1} {i} \rb
\label{eq:band matrix 2}\\
&  \leq  - \f 12 \log 1 =0 \nonumber,
}
where \eqref{eq:band matrix 2} follows from properties of symmetric tri-diagonal matrices.
With the bounding in  \eqref{eq:band matrix 2}, we obtain the outer bound
\ea{
R^{\rm OUT}
& \leq \f 1 2 \log(1+P+c^2) - \f {M-1} {2M} \log c^2 + \f 3 2.
\label{eq:outer M CCDP}
}
As for the expression in \eqref{eq:out independent 1}, the outer bound in \eqref{eq:outer M CCDP} is convex in $c^2$ with  a minimum in
\ea{
{c^2}^*= (M-1)(P+1).
}
Using Lem. \ref{lem:capacity decreasing scaling fading} to substitute $c^2$ with  $\min \{c^2, (M-1)(P+1)\}$ in the expression of
 \eqref{eq:outer M CCDP}, together with some further bounding, yields the outer bound  in \eqref{eq: outer bound CCDP bernoulli}.

\smallskip
\noindent
$\bullet$ \emph{Achievability:}
The value of   $\alb$ which maximizes \eqref{eq:al opt M} is
\ea{
\alb^*= \max \lcb 0,\min \lcb 1, \f{c^2+1-M}{P(M-1)} \rcb \rcb,
\label{eq:al opt M}
}
and the above scheme reduces to simple time-sharing and state pre-cancellation when $c^2>(M-1)(P+1)$.
Using the optimal power allocation in \eqref{eq:al opt M}, we obtain the inner bound
\ea{
& R^{\rm IN} = \label{eq:inner bound strong} \\
& \lcb \p{
 \f 12 \log \lb 1 + \f{P} {1+c^2} \rb & M-1 >  c^2 \\
\f 12 \log (P+c^2  +1)   &  {M-1}  \leq  c^2  \leq (M-1) (P+1) \\
\quad - \f {M-1}{2 M } \log \lb c^2 \rb-\f 12  \\
 \f {1} {2M} \log(1+P)     & c^2> (M-1)(P+1).
}\rnone \nonumber
}

\smallskip
\noindent
$\bullet$ \emph{Gap to capacity:}
%
Compare the expression in  \eqref{eq:outer M CCDP} and in \eqref{eq:inner bound strong} for $M>2$:
the largest gap between inner and outer bound is $2.25 \ \bpcu$ and is attained for $M-1 \leq c^2$.
In all other regimes is at most $2 \ \bpcu$.

\section{Proof of Th. \ref{th:Outer bound strong}.}
\label{app:Outer bound strong}
The derivation of the outer bound involves extending the bounding in Th. \ref{th:Approximate capacity for the $2$-receiver  WSFD}
in the strong fading regime
%
%
to the case of any number of possible fading realization.
The key in the derivation is a careful choice of the genie-aided side information provided at each compound receiver.

\noindent
\noindent
\emph{Converse:}
The derivation employs a recursion involving a number of algebraic derivations: we first
establishing this recursion for $M=3$, then consider the case of any $M$.

\smallskip
\noindent
$\bullet$ {\emph{Case for $M=3$:}}
Consider a $3$-receiver WSFD channel for which the conditions in \eqref{eq: strong fading} hold, then
\eas{
&  I(Y_1^N;W)+I(Y_2^N;W) \nonumber \\
& = H(Y_1^N)+H(Y_2^N)-H(c a_2 S^N +\Zt_2^N,Y_1^N|W)
\label{eq:pp1}\\
& = H(Y_1^N)+H(Y_2^N)-H(c a_2 S^N +\Zt_2^N) \nonumber \\
& \quad \quad  -H(Y_1^N|W,c a_2 S^N +\Zt_2^N)
\label{eq:pp2}\\
& = \f N 2 \log(1+P)+\f N 2 \log \lb 1+ \f {P} {1+c^2 a_2^2}\rb \nonumber \\
& \quad \quad -H(Y_1^N|W,c a_2 S^N +\Zt_2^N)+2N+N \log  2 \pi e,
\label{eq:pp3}
}{\label{eq:ppppp}}
where \eqref{eq:pp1} follows from the assumption that $a_1=0$ and by letting $\Zt_2^2=Z_2^N-Z_1^N$.

Using Fano's inequality, the capacity can be bounded as
\eas{
& 3N (R-\ep) \nonumber \\
& \leq    I(Y_1^N;W)+I(Y_2^N;W)+I(Y_3^N;W)  \nonumber \\
& \leq I(Y_1^N;W)+I(Y_2^N;W) \nonumber \\
& \quad \quad +I(Y_3^N,c a_2 S^N +\Zt_2^N;W) \label{eq:pp4} \\
& = I(Y_1^N;W)+I(Y_2^N;W) \nonumber \\
 & \quad \quad +I(Y_3^N;W|c a_2 S^N +\Zt_2^N)
\label{eq:pp5}\\
& = \f N 2 \log(1+P)+\f N 2 \log \lb 1 + \f {P} {1+c^2 a_2^2}\rb \nonumber \\
& \quad \quad + H(Y_3^N|c a_2 S^N +\Zt_2^N) \nonumber \\
& \quad \quad  -\lb H(Y_3^N|W,c a_2 S^N +\Zt_1^N) -H(Y_1^N|W,c a_2 S^N +\Zt_2^N) \rb \nonumber\\
& \quad \quad   +2N+ N \log (2 \pi e)
\label{eq:pp6} \\
& = \f N 2 \log(1+P)+\f N 2 \log \lb 1 +\f {P} {1+c^2 a_2}\rb \nonumber \\
& \quad \quad + H(Y_3^N|c a_2 S^N +\Zt_2^N) \nonumber \\
& \quad \quad - \lb H(Y_3^N,Y_1^N|W,c a_2 S^N +\Zt_2^N)\rb \nonumber \\
& \quad \quad +2N+ N \log (2 \pi e),
\label{eq:pp7}
}{\label{eq:pppp1}}
where \eqref{eq:pp4} follows by providing $c a_2 S^N +\Zt_2^N$ as a side information to the third compound receiver and
\eqref{eq:pp5} follows from the independence of the message from the channel state and \eqref{eq:pp6} from \eqref{eq:ppppp}.

Continuing the series of inequalities in \eqref{eq:pppp1}:
\eas{
\eqref{eq:pp7}
& \leq \f N 2 \log(1+P)+\f N 2 \log \lb 1+ \f {P} {1+c^2 a_2}\rb \nonumber \\
& \quad \quad + H(Y_3^N|c a_2 S^N +\Zt_2^N) \nonumber  \\
& \quad \quad  - H(c a_3 S^N + \Zt_3^N |c a_2 S^N +\Zt_3^N) \nonumber \\
& \quad \quad + H(Z_1^N|Z_3^N-Z_1^N,Z_2^N-Z_1^N)  \nonumber\\
& \quad \quad  +2 N + N \log (2 \pi e)
\label{eq:pp8} \\
& \leq \f N 2 \log(1+P)+\f N 2 \log \lb 1+ \f {P} {1+c^2 a_2}\rb \nonumber \\
& \quad \quad + H(Y_3^N|c a_2 S^N +\Zt_2^N) \nonumber \\
& \quad \quad   -N  H(c a_3 S_i + \Zt_{3i} |c a_2 S_i +\Zt_{3i}) \nonumber \\
& \quad \quad +  N H(Z_{1i}|Z_{3i}-Z_{1i},Z_{2i}-Z_{1i}) \nonumber\\
& \quad \quad  +2N + N \log (2 \pi e),
\label{eq:pp9}
}{\label{eq:m=3 passage 1}}
where \eqref{eq:pp8} follows from letting  $\Zt_3^N=Z_3^N-Z_1^N$.
The entropy term  $H(Y_3^N|c a_2 S^N +\Zt_2^N)$  in \eqref{eq:pp9} can be bounded using the conditional version of the GME property as follows
\ea{
& H(Y_3^N|c a_2 S^N +\Zt_2^N)  \nonumber \\
& \leq \max_{\rho_{XS}} \f  N 2 \log \lb P + c^2 a_3^2 + 2 c a_3 \rho_{PS} \sqrt{P}+1 \rnone \nonumber \\
& \quad \quad  \quad \quad  \lnone - \f {c^2 a_2^2} { c a_2^2 +2} \lb \rho_{XS} \sqrt{P}+c a_3 \rb^2  \rb \nonumber \\
& \quad \quad + \f N2 \log(2 \pi e),
\label{eq:max rhoXS}
}
where the covariance between the \emph{i.i.d.} Gaussian version of $X$ and $S$ is
$\rho_{XS}\sqrt{P}$.
The expression in \eqref{eq:max rhoXS} attains a maximum in $\rho_{XS}$ for
\ea{
\rho_{XS}^*=\f {2 a_3}{c \sqrt{P} a_2^2},
\label{eq:max value rho_{XS}}
}
yielding the bound
\ea{
& H(Y_3^N|c a_2 S^N +\Zt_2^N)  \nonumber \\
& \leq \f N 2 \log \lb P+1+2 \f {a_3^2} {a_2^2} \rb+\f N2 \log(2 \pi e).
\label{eq:bound positive entropy M strong}
}

By evaluating the entropy expressions  $H(a_3 S_i + \Zt_{3i} |c a_2 S_i +\Zt_{2i})$ and $ H(Z_{1i}|Z_{3i}-Z_{1i},Z_{2i}-Z_{1i})$,
we finally come to the outer bound
\ea{
& 3N (R-\ep) \leq \f N 2 \log(1+P)+\f N2 \log \lb \f {1+P} {1+c^2 a_2^2}\rb \nonumber \\
& \quad \quad \quad \quad  +\f N 2 \log \lb \f { c^2 a_2^2+2}{c^2 a_2^2} \f {(P+1) a_2^2+2 a_3^2}{a_3^2+a_2^2}  \rb \nonumber \\
& \quad \quad \quad \quad  - \f N 2 \log \lb \f 1 3 \rb.
\label{eq:some stuff}
}
Since $c^2 a_2^2 \geq P \geq 1$ and $a_3^2 \geq P a_2^2$, we have that
\ea{
& \f { c^2 a_2^2+2}{c^2 a_2^2} \f {(P+1) a_2^2+2 a_3^2}{a_3^2+a_2^2}  \geq \f 9 2,
}
and moreover
\ea{
\f 1 2 \log \lb 1+ \f {P} {1+c^2 a_2}\rb \leq \f 12,
}
so that we obtain the outer bound
\ea{
R^{\rm OUT}
& = \f 1 3 \lb \f 1 2 \log(P+1) + \f 12  + \f  1 2  \log \f 9 2 + \f 1 2 \log 3 \rb \nonumber \\
& = \f 1 6 \log (P+1) + 1.
\label{eq:outer 3}
}

\smallskip
\noindent
$\bullet$ {\emph{Case for a general $M$}:}
Next we wish to generalize the derivation in \eqref{eq:outer 3} to the case of any $M$:
this can be accomplished by providing each user with the appropriate side information and employing a recursion as in the converse proof in Th. \ref{th:Approximate capacity $M$-CCDP with Gaussian independent states}.

We begin by bounding the capacity as in \eqref{eq:label M case} to obtain
\ea{
& N(R-\ep) \leq  \f 1 M \sum_{m=1}^M \f N 2 \log (P+c^2a_m^2+2 c a_m^2 \sqrt{P}+1) \nonumber \\
& \quad \quad +\f N {2 M} \log (2 \pi e) - \f 1 M  \sum_{m=1}^M  H(Y_m^N|W).
\label{eq:fano wsfd}
}
Next, we define
\eas{
\Zt_m^N & =Z_{m}^N-Z_1^N \\
V_2^N & = \emptyset \\
V_m^N &= [ V_{m-1}^N, \ a_{m-1} S^N +\Zt_{m-1}^N],
}{\label{eq:def 2}}
for $m>1$.
Using the definitions in \eqref{eq:def 2}, we continue the bounding in \eqref{eq:fano wsfd} as
\eas{
\eqref{eq:fano wsfd}
& \leq  \f 1 M \sum_{m=1}^M I(Y_m^N, V_m^N ;W) \\
& \leq  \f 1 M \sum_{m=1}^M I(Y_m^N ;W | V_m^N),
\label{eq:m=3 passage 1}
}{\label{eq:m=3 passage}}
where \eqref{eq:m=3 passage 1} follows from the fact that the state sequences are independent from the message $W$.

Let $K_m$ be defined as
\ea{
K_m = \sum_{j=m}^{M} I(Y_j^N,W| V_j^N),
\label{eq:def Tm wsfd}
}
and rewrite \eqref{eq:m=3 passage} as
\eas{
& M (N R-\ep) \nonumber \\
&  \leq
I(Y_1^N;W)+I(Y_2^N;W)+ K_3 \\
& \leq  \f N 2 \log2 \pi e (1+  P)+\f N 2 \log 2 \pi e \lb1 + \f {P} {1+c^2 a_2^2}\rb  \nonumber \\
& \quad \quad -H(Y_1^N|W,c a_2 S^N +\Zt_2^N) + K_3
\label{eq:strong fano 1}\\
& \leq  \f N 2 \log(1+  P)+\f N 2 \log \lb 1+ \f {P} {1+c^2 a_2^2}\rb  + N \log (2 \pi e) \nonumber \\
& \quad \quad -H(Y_1^N|W,c a_2 S^N +\Zt_2^N) + I(Y_3^N;W|V_3^N)  \nonumber \\
& \quad \quad \quad \quad + K_4,
}{\label{eq:strong fano}}
where \eqref{eq:strong fano 1}
follows from the bound in \eqref{eq:pppp1}.
With a derivation similar to \eqref{eq:some stuff} and given the conditions in \eqref{eq: strong fading}, we obtain
\eas{
& K_4-H(Y_1^N|W,c a_2 S^N +\Zt_2^N) + I(Y_3^N;W|V_3^N)  \nonumber \\
& \leq K_4 -H(Y_1^N|W,c a_2 S^N +\Zt_2^N, c a_3 S^N+\Zt_3^N) \nonumber \\
& \quad \quad + \f N 2 \log(9/ 2) \\
& = K_4 -H(Y_1^N|W,V_4^N)+ \f N 2 \log(9/ 2).
}{\label{eq:bonding first step}}
The bounding in \eqref{eq:bonding first step} can be recursively repeated as
\eas{
& K_m -H(Y_1^N|W,V_m^N) \nonumber  \\
&  = K_{m+1}-H(Y_1^N|W,V_m^N) + I(Y_m^N;W|V_m^N)  \\
& \leq  K_{m+1}-H(Y_1^N|W,V_{m+1}^N)+ \ka_m,
}{\label{eq:recursion M}}
for $\ka_m$ defined as
\ea{
\ka_m & = H(Y_m^N|V_m^N)- H(c a_m S^N+\Zt_m| W,V_m^N) \nonumber \\
    & = H(Y_m^N|V_m^N)- H(V_{m+1}^N| V_m^N).
\label{eq:k_m}
}
By repeating the recursion step in \eqref{eq:recursion M} $M-2$ times, we come to the outer bound
\ea{
M (N R-\ep)
& \leq \f 1 2 \log2 \pi e (P+1) \nonumber\\
& \quad + \f 12 \log2 \pi e  \lb1+ \f{P}{c^2 a_2^2+1} \rb \nonumber \\
& \quad + \sum_{m=2}^M k_m -  H(Y_1^N|W,V_{M+1}^N).
\label{eq:strong fano M}
}
We next wish to show that the terms $\ka_m$ and $H(Y_1^N|W,V_{M+1})$ in the RHS of \eqref{eq:strong fano M} are bounded by a
 constant for all parameter regimes and for a given value $M$.

Let's begin by bounding the term $H(Y_m|V_m)$ in \eqref{eq:k_m}:
{\small
\ea{
& H(Y_m^N|V_m^N) \nonumber \\
& = H( Y_m^N| V_m^N,Z_1^N) +I(Y_m^N;Z_1^N|V_m^N) \nonumber \\
& \leq H(Y_m^N | c a_2 S^N +Z_2^N, c a_3 S^N + Z_3^N \ldots c a_{m-1} S^N + Z_{m-1}^N,Z_1^N) \nonumber \\
& \quad \quad + H(Z_1^N)-H(Z_1^N | V_m^N,Y_m^N ) \nonumber \\
&  \leq H(Y_m^N| c a_2 S^N +Z_2^N, c a_3 S^N + Z_3^N \ldots c a_{m-1} S^N + Z_{m-1}^N) \nonumber \\
& \quad \quad + H(Z_1^N) -H(Z_1^N | \Zt_2^N, \Zt_3^N \ldots \Zt_{i-1}^N,S^N) \nonumber \\
%
&  \leq H(Y_m^N | c a_2 S^N +Z_2^N, c a_3 S^N + Z_3^N \ldots c a_{i-1} S^N + Z_{m-1}^N) \nonumber \\
& \quad \quad -H \lb Z_1^N | Z_1^N + \f { \sum_{i=2}^{m-1}Z_i }{q-2} \rb \nonumber \\
& \quad \quad  + \f N 2 \log(2 \pi e) \nonumber  \\
&  \leq H(Y_m^N | c a_2 S^N +Z_2^N, c a_3 S^N + Z_3^N \ldots c a_{i-1} S^N + Z_{m-1}^N) \nonumber \\
& \quad \quad  + \f N 2 \log(m-1).
\label{eq:k p 1}
}
}
Consider the term $H(Y_m^N | c a_2 S^N +Z_2^N \ldots c a_{m-1} S^N + Z_{m-1}^N)$: the random variables in the conditioning are noisy version of $S^N$
and thus a sufficient statistic can be obtained by applying the maximal ratio combining principle.
This yields the estimate $S^N+\Zh_i$  of $S^N$ for
\ea{
 \Zh_m =  \f {\sum_{j=2}^{m-1} c a_j Z_j}{\sum_{j=2}^{m-1} c^2 a_j^2}  \sim \Ncal (0,\widehat{\sigma}^2),
}
and
\ea{
\widehat{\sigma}^2 =  \f {1} {\sum_{j=2}^{m-1} c^2 a_j^2},
}
so that
\ea{
& H(Y_m^N| c a_2 S^N +Z_2^N \ldots c a_{m-1} S^N + Z_{m-1}^N) \nonumber \\
& = H(Y_m^N | S+\Zh_m).
\label{eq:bound sufficient statistic strong}
}
As for the bounding in \eqref{eq:bound positive entropy M strong}, we have that \eqref{eq:bound sufficient statistic strong}
can be bounded using the GME property and  by optimizing over the correlation between the Gaussian version $S_m$ and $X_m$.
Again through the GME property, we obtain the outer bound
\ea{
& H(Y_m^N | c a_2 S^N +Z_2^N \ldots c a_{m-1} S^N + Z_{m-1}^N) \nonumber \\
&  \leq  \f N 2 \log 2 \pi e\lb 1+ P + \f {a_m^2} {\sum_{j=2}^{m-1}  a_j^2 } \rb,
\label{eq:k p 2}
}
where, in \eqref{eq:k p 2}, we have used again an optimization similar to \eqref{eq:max value rho_{XS}} yielding the optimal correlation
\ea{
\rho_{XS}^* =\f { c a_m \widehat{\sigma}^2} {\sqrt{P}}.
}
Next, we bound the term $H(V_{m+1}^N| V_m^N)$ in \eqref{eq:k_m}:
\ea{
&H(V_{m+1}^N| V_m^N) \nonumber  \\
& \leq -H(V_{m+1}^N| V_m^N,Z_1^N) \nonumber \\
& \leq -H(c a_m S^N+Z_i^N| c a_2 S +Z_2^N \ldots c a_{m-1} S^N +Z_{m-1}^N) \nonumber \\
& = -H(c a_m S^N+Z_i^N|S^N +\Zh^N) \nonumber \\
& =  \f N 2 \log  2 \pi e  \lb c^2 a_m^2 +1 - \f {c^2 a_m^2}{1 + \widehat{\sigma} ^2 } \rb   \nonumber \\
& =   \f N 2 \log 2 \pi e  \lb c^2 a_m^2 +1 -\f 1 {1 + \sum_{j=2}^{m-1} c^2 a_j^2 } \rb   \nonumber \\
&  = \f N 2 \log 2 \pi e  \lb  \f {1+ \sum_{j=2}^{m} c^2 a_j^2} {1 + \sum_{q=2}^{m-1} c^2 a_q^2 } \rb.
\label{eq:k p 3}
}
Combining the results in \eqref{eq:k p 1}, \eqref{eq:k p 2} and \eqref{eq:k p 3}, we can bound  the $\ka_m$ in \eqref{eq:recursion M} as
\eas{
\ka_m
& =  \f N2 \log \lb 1+ P + \f {c^2 a_m^2} {\sum_{j=2}^{m-1} c^2 a_j^2 } \rb + \f N2 \log(m-1)   \nonumber \\
& \quad \quad -  \f N2 \log \lb  \f {1+ \sum_{j=2}^{m} c^2 a_j^2} {1 + \sum_{j=2}^{m-1} c^2 a_j^2 } \rb- \f N 2 \log (2 \pi e) \\
& = \f N2 \log \lb
\lb \f {\sum_{j=2}^{m-1} c^2 a_j^2 +1}{\sum_{j=2}^{m-1} c^2 a_j^2 }  \rb  \cdot \rnone  \\
& \quad \quad \lnone \lb \f {(P+1) \sum_{j=2}^{m-1} c^2 a_j^2  + c^2 a_m^2 } { \sum_{j=2}^{m} c^2 a_j^2  }  \rb
\rb + \f N2 \log(m-1). \nonumber
}{\label{eq:constant c}}
Using the conditions in \eqref{eq: strong fading}, we have that the bound in \eqref{eq:constant c} can be further loosened as
\ea{
\ka_m \leq  N \lb  2 + \f 1 2 \log(m-1)\rb,
}
so that
\ea{
\f 1 N \sum_{i=2}^M k_i \leq M + \f 1 2  \log \f {M!} M  \leq \f {M-1} 2 \log M +2M.
\label{eq:k bound}
}
The term  $H(Y_1^N|W,V_{M+1}^N)$ in \eqref{eq:strong fano M} can be bounded as:
\ea{
& -H(Y_1^N|W,V_{M+1}^N)\leq -H(Y_1^N|W,V_{M+1}^N, X^N,S^N) \nonumber \\
& =\f N2 \log  2 \pi e M.
\label{eq:boud Y1}
}
Substituting the bounds in \eqref{eq:boud Y1} and \eqref{eq:k bound} in \eqref{eq:strong fano M} we obtain the expression in \eqref{eq:outer bound strong}.

\smallskip
\noindent
$\bullet$ {\bf Capacity inner bound and approximate capacity:}
Consider the scheme in which the encoder transmits to the $m^{\rm th}$ compound receiver as in the WDP channel for a portion $1/M$ of the time:
this scheme attains
\ea{
R^{\rm IN}= \f 1 {2M} \log(1+P).
\label{eq:time sharing}
}
The gap between inner bound in \eqref{eq:time sharing} and outer bound in  \eqref{eq:outer bound strong} is
\ean{
& \f 1 M \lb \f 12 \log2 \pi e  \lb1+ \f{P}{c^2 a_2^2+1} \rb + \sum_{m=2}^M \ka_m \rnone \nonumber \\
&  \quad \quad  \lnone -  \f 1 N H(Y_1^N|W,V_{M+1}^N)-\f 1 2\log(2\pi e) \rb \leq \f 12 \log M+2.
}

\section{Proof of Lem. \ref{lem:Outer bound strong v2}}
\label{app:Outer bound strong v2}

This Lemma is shown by
adapting the derivation of the outer bound in Th. \ref{th:Outer bound strong} : the inner bound derivation is not affected by
these more general conditions.

\smallskip
\noindent
$\bullet$ {\bf Capacity outer bound:}
If $a_1 \neq 0$, \eqref{eq:strong fano M} becomes
\ea{
& R^{\rm OUT}  = \f 1 2 \log(P+1+c^2a_1^2) + \f 12 \log\lb \f{P+c^2 a_2^2+1}{c^2 (a_2-a_1)^2+1} \rb  \nonumber\\
& \quad + \sum_{m=2}^M \ka_m  + \sum_{m=2}^M -H(Y_1|c a_2 S^N+\Zt_2 \ldots c a_M S^N+\Zt_M),
}
while \eqref{eq:constant c} becomes
\eas{
\ka_m & = \f 12 \log \lb 1+ P + \f {c^2 a_m^2} {\sum_{j=2}^{m-1}  c^2 \Delta_j^2 } \rb + \f 1 2 \log(m-1) \nonumber \\
& \quad \quad -  \f 1 2 \log \lb  \f {1+ \sum_{j=2}^{m} c^2 \Delta_j^2} {1 + \sum_{j=2}^{m-1} c^2 \Delta_j^2 } \rb \\
 & = \f 1 2 \log \lb \f {1 + \sum_{j=2}^{m-1} c^2 \Delta_j^2}{\sum_{j=2}^{m-1} c^2 \Delta_j^2} \cdot  \rnone \nonumber \\
 & \quad \quad \lnone \lb \f{(P+1)  \sum_{j=2}^{m-1} c^2 \Delta_j^2 } {1 + \sum_{j=2}^{m} c^2 \Delta_j^2} + \f {c^2 a_m^2}
 {1+ \sum_{j=2}^{m} c^2 \Delta_j^2} \rb  \rb \nonumber \\
& \leq 2 + \f 1 2 \log(\gamma) + \f 12 \log(m).
}
From conditions on $a_2$ and $\Delta_2$ in \eqref{eq: strong fading v2}, we come to the outer bound in \eqref{eq:outer bound strong v2}.

The inner bound  derivation is not affected by the assumption that $a_1 \neq 0$ and thus the gap from capacity is adjusted by adding the term $1/2\log(\gamma)$.

\section{Proof of Th. \ref{th:Approximate capacity for the  $2$-receiver  CCDP correlated}}
\label{app:Approximate capacity for the  $2$-receiver  CCDP correlated}

\smallskip
\noindent
$\bullet$ {\bf Capacity outer bound:}
Consider the outer bound in \eqref{eq: outer bound CCDP bernoulli} obtained by providing $S_c$ to both decoders. The receivers can remove this sequence from the channel output: the corresponding output is the same model as in
Th. \ref{th:Approximate capacity for the  $2$-receiver  CCDP with Gaussian independent states} but with a state with smaller variance, that is $1-\rho$ instead of $\rho$.
By absorbing this factor in $c$, we obtain the outer bound in \eqref{eq: outer bound $2$-receiver  CCDP correlated}.

\smallskip
\noindent
$\bullet$ {\bf Capacity inner bound and approximate capacity:}
The inner bound for this scenario is again an extension of the inner bound in Fig. \ref{fig:achievableSchemeSuperposition} with the difference that the base codeword now pre-codes against
the sequence $S_c^N$:
the attainable rate for each user is
\ean{
R^{\rm IN} = \f 12 \log \lb 1+ \f {\al P }  {1 + \alb P + c^2(1-\rho)}\rb + \f 1 4 \log (1+ \alb P).
}
By optimizing over the parameter $\al$ we obtain the inner bound
\ea{
R^{\rm IN} = \lcb \p{
\f 12 \log \lb 1+ \f P {c^2 (1-\rho) +1}\rb               & c^2(1-\rho) < 1 \\
\f 12 \log \lb 1+ c^2(1-\rho) + P\rb \\
\quad \quad -\f 14 \log(c^2(1-\rho))- \f 12   & 1 \leq c^2 (1-\rho) < P+1\\
\f 14 \log (P+1)               & c^2 (1-\rho) \geq P+1.
}\rnone
\label{eq:last eq}
}
The expression in \eqref{eq:last eq} is to within $1 \ \bpcu$ from the outer bound in \eqref{eq: outer bound $2$-receiver  CCDP correlated}.

%
%
%
%
%
%

\end{document}